\documentclass[fleqn,usenatbib]{mnras}

\usepackage{newtxtext,newtxmath}

\usepackage[T1]{fontenc}
\usepackage{ae,aecompl}

\usepackage{graphicx}	
\usepackage{amsmath}	
\usepackage{amssymb}	
\usepackage[export]{adjustbox}
\usepackage[usenames,dvipsnames]{xcolor}
\usepackage{natbib}
\usepackage{multirow}
\usepackage{booktabs}
\usepackage[flushleft]{threeparttable}

\newcommand*{\rom}[1]{\expandafter\@slowromancap\romannumeral #1@}
\newcommand{\ie}{\textit{i}.\textit{e}., }
\newcommand{\eg}{\textit{e}.\textit{g}. }
\makeatother

\title[xgb Stellar classification]{General classification of light curves using extreme boosting}

\author[R. Kgoadi et al.]{Refilwe Kgoadi,$^{1}$\thanks{E-mail: refilwe.kgoadi1@my.jcu.edu.au (RK)}
Chris Engelbrecht,$^{2}$
Ian Whittingham$^{1}$
and Andrew Tkachenko$^{3}$
\\
$^{1}$College of Science and Engineering, James Cook University, Townsville, Australia, 4811\\
$^{2}$Physics Department, Faculty of Science, University of Johannesburg, South Africa, 2006\\
$^{3}$Instituut voor Sterrenkunde, KU Leuven, Celestijnenlaan 200D, B-3001 Leuven, Belgium
}

\date{Accepted XXX. Received YYY; in original form ZZZ}

\pubyear{2019}

\begin{document}
\label{firstpage}
\pagerange{\pageref{firstpage}--\pageref{lastpage}}
\maketitle

\begin{abstract}
A significant degree of misclassification of variable stars through the application of machine learning methods to survey data motivates a search for more reliable and accurate machine learning procedures, especially in light of the very large data cubes that will be generated by future surveys and the need for immediate production of accurate, formalised catalogues of variable behaviour to enable science to proceed. In this study, the efficiency of an ensemble machine learning procedure utilising extreme boosting was determined by application to a large sample of data from the OGLE III and IV surveys and from the \textit{Kepler} mission. Through recursive training of classifiers, the study developed a variable star classification workflow which produced an average efficiency determined with the average precision of the model (0.81 for \textit{Kepler} and 0.91 for OGLE) and the $f-score$ of predictions on the test sets. This suggests that extreme boosting can be presented as one of the favourable shallow learning methods in developing a variable star classifier for future large survey projects.
\end{abstract}




\section{Introduction}\label{Sec:intro}

High-precision photometry from satellite missions such as the CoRoT \citep{Michel2008}, MOST \citep{Rucinski2003,Barban2007} and \textit{Kepler} \citep{Koch2010} missions has made huge contributions to asteroseismology. These include asteroseismology across a wide temperature range \citep{Gilliland2010, Chaplin2014}, a regularly updated catalogue of eclipsing binaries in the \textit{Kepler} field initiated by \citet{Prsa2011} and characterisation of stellar rotation \citep{Reinhold2013}. Significant developments include the revision of the asteroseismological Hertzsprung-Russell (H-R) Diagram and the confirmation of the existence of both pressure (p-) and gravity (g-) mode frequency regimes in $\delta$ Scuti and $\gamma$ Dor stars \citep{Grigahcene2010}. These modes aid in probing physical conditions in the stellar envelope and stellar interior, providing information on the internal dynamics of a star. Furthermore, a study by \citet{Casanellas2010} shows that it is plausible to detect dark matter through asteroseismology. This was reinforced by \citet{Martins2017} who argue that asteroseismology of solar-like stars may be used as a complementary tool in understanding properties of dark matter.

Although space missions have transformed the science of stellar physics, formalised catalogues of observed objects from large scale missions do not exist. Preliminary studies towards formalised catalogues for the CoRoT and \textit{Kepler} missions through automation have been conducted by \citet{Debosscher2009,Debosscher2011} for example. Furthermore, these classifiers were supplemented by classification of stars on an individual basis where stars were studied within a specific temperature range. For example, \citet{Balona2011e,Balona2015} focused on hot stars whilst cooler stars were studied by \citet{Uytterhoeven2011} and an evolving catalogue of eclipsing binaries has been generated by \citet{Kirk2016} . The sheer number of observations produced by current and future survey missions makes human classification a practical impossibility. As no automated procedure can be perfect when applied to a vast number of light curves of varying quality, there will always remain a proportion of misclassified stars. The challenge for automated classification is to bring this proportion down as low as possible. 

It would be ideal if large scale survey operations could be subjected to the most accurate automated classification procedures possible, so that reliable formalised catalogues could be produced. Studies by \citet{Blomme2011} and \citet{Sarro2013} suggest that automated classifiers similar to those of \citet{Eyer2005} and \citet{Debosscher2009,Debosscher2011} can be improved with the help of Knowledge Discovery Processes (KDPs). These processes occur when machine learning algorithms are applied to a plethora of stellar data from current missions such as the Transiting Exo-planet Survey Satellite \citep[TESS,][]{Ricker2014} and imminent ones like the James Webb Space Telescope \citep[JWST,][]{Beichman2014} and PLAnetary Transits and Oscillations of stars \citep[PLATO,][]{Rauer2014}. Early stage classification will not only aid with census studies, it will catalyse in-depth scientific studies that are needed to understand our universe.

A review by \citet{Djorgovski2013} on sky surveys has emphasised the magnitude and rate at which astronomical data are increasing, and \citet{Ball2009} and \citet{Borne2013} have shown that KDPs can be applied to astronomical data with minimal intervention. This was further validated by \citet{Oza2012}, \citet{Wagstaff2012} and \citet{Lin2012} through the classification reviews summarised in table \ref{Tab:stellarclassificationML}. An Astroinformatics review by \citet{Borne2009} focuses on the use of Data Science as a tool in astronomical data analysis. 

\begin{table*}
\scriptsize
\caption{Shallow and deep classifiers applied to stellar light curves }\label{Tab:stellarclassificationML} 
\centering 
\begin{tabular}{l l p{9.5cm} } 
\hline
Method & Technique & Example \\[0.5ex]
\hline
\multirow{4}{*}{Supervised} & Support Vector Machines (SVMs) & Classification of sparse and noisy light curves \citep{Debosscher2009,Richards2011}. \\[0.6ex]
& Classification trees & Hierarchical classification of variable stars \citep{Silla2011}. \\[0.6ex]
& Random Forests (RFs) & K2 and ASAS-SN catalogues of variable stars \citep{Armstrong2016,Jayasinghe2018}\\[0.6ex]
 & Artificial Neural Networks (ANNs) & Application of Neural Networks to classify stars \citep{Sarro2006,Feeney2005}. \\[0.6ex]
 \hline
\multirow{2}{*}{Unsupervised} & Gaussian Mixture Models & Clustering of variable stars using Expectation Maximisation \citep{Eyer2005, Sarro2009b}. \\[0.6ex]
& Self Organising Maps (SOM) & Application of SOM parameters to classify stars \citep{Brett2004, Graczyk2011}. \\[0.6ex] 
\hline
\multirow{2}{*}{Deep Learning} & Convolution Neural Network (CNN) & Classification of image data generated from Catalina Real-time Transient Survey \citep[CRTS,][]{Graczyk2011} and \textit{Kepler} \citep{Hon2018b} light curves using CNN. \\[0.6ex]
 & Classification in asteroseismology & Analysis of power spectra of oscillating red giants \citep{Hon2017, Hon2018}. \\[0.9ex]
\hline
\end{tabular}
\end{table*}

Even though the methods listed in table \ref{Tab:stellarclassificationML} have improved how data are analysed in astronomy, ensemble Random Forests (RFs) \citep{Breiman2001, Kim2016, Armstrong2016, Jayasinghe2018, Jayasinghe2018a} and hybrid methods such as Modal Expectation Maximisation \citep{Li2007, Sarro2009b} have proven to be more robust and effective than traditional machine learning (ML) methods with regard to classifying variable stars. This paper will explore one of the rapidly evolving ensemble methods that is based on gradient boosted trees \citep{Freidman2001, Friedman2002, Natekin2013}, applied through extreme boosting \citep{Chen2016}. For the purpose of the paper, this model will be named as the \verb"xgb" classifier applied to features (input variables). This paper aims to introduce extreme gradient boosted trees as an alternative classifier for variable stellar light curves by applying it to high-dimensional data-frames derived from Optical Gravitational Lensing Experiment \citep[OGLE,][]{Udalski2008, Udalski2015} and \textit{Kepler} mission \citep{Koch2010} data. The development of the workflow associated with the application of ML algorithms to stellar light curves will be discussed in sections \ref{Sec:Data}, \ref{Sec:period} and \ref{Sec:workflow}. This includes acquisition and rejection of the analysed data (section \ref{Sec:Data}), an optimised period searching method (section \ref{Sec:period}), derivation of features from light curves using a combination of methods, selection of features used to optimise the classifier and the efficiency of the classifier (section \ref{Sec:workflow}). The strength and weakness of the classifier and its application to the \textit{Kepler} and OGLE and data sets will be discussed in section \ref{Sec:results}.

\section{Data}\label{Sec:Data}

\subsection{Acquisition}\label{Sec:acquisition}

Data used in this paper were derived primarily from stellar light curves. Supplementary data such as colour indices of stars in the data-frames were sourced from respective survey archived data. In the case of \textit{Kepler} data, two sets of colour indices were used. These were $J - K$ from the \textit{Kepler} archive and $G_{bp} - G_{rp}$ sourced from \citet{Bailer2018}. Addition of colour indices was done to enhance separation of stars with overlapping variability properties. Given that this paper focuses on a classification method, already labelled stellar light curves for the above mentioned data-sets were used. Labels for each of the data-sets will be discussed in the sub-sections below. 

\subsection{OGLE data-set} 
 
The Optical Gravitational Lensing Experiment is one of the longest running ground-based sky surveying projects to date. Its first operation was in 1992 \citep{Udalski2015} with first results released in 1994 \citep{Udalski1994}. The OGLE projects mainly observed the Magellanic System and the Galactic Bulge, focusing on the discovery of celestial objects and sky variability studies. To date there have been four successive projects at the Las Campanas Observatory in Chile and the latest data release of OGLE-IV with updates of previously observed objects was in July 2017.

Variable stars are amongst other objects of interest that have been observed and analysed, with over 200,000 detected. Projects such as OGLE are crucial to time domain astronomy since they aid in retaining historical information that can be used in prediction studies. Such studies can then be used as a case study to show that long scale studies are valuable in reclassification based on more input data. One such example is by \citet{Soszynski2015a}, where a number of stars that were previously classified as Classical Cepheids from OGLE III are now categorised as other types of variables. This was done by comparing Fourier parameters of light curves. Such long-term studies can also assist in validation and/or modification of theoretical astrophysical models.

Light curves obtained through image differencing data used in this paper are publicly accessible from the OGLE website \footnote{\url{http://ogle.astrouw.edu.pl/}}. For this paper, I-filter light curves of variable stars in the Large Magellanic Cloud (LMC) were used and labels for stars were sourced from the OGLE website. Periods estimated here were compared to the ones published by OGLE.

Classes for this data-set are from the OGLE III \citep{Udalski2008} and IV \citep{Udalski2015} data-sets. The data-set consisted of five main types of variables in the LMC field. These are classical Cepheids \citep{Soszynski2015a}, $\delta$ Scuti stars \citep{Poleski2010}, eclipsing binaries and ellipsoidal variables \citep{Pawlak2016}, Long Period Variables (LPVs) \citep{Soszynski2009} and RR Lyrae \citep{Soszynski2016} stars. Cepheids were categorised into two groups; Cepheids and Cepheids$\_$F. The former was a group of stars with 1O or 2O pulsation modes whereas the latter have F modes. LPVs consisted of three classes of stars with either longer or semi regular variability trends. These were Miras, semi regular variables (SRVs) and OGLE small amplitude red giants (OSARG) stars. RR Lyrae stars were categorised as RR \textit{ab} and RR \textit{Lyrae} stars. The latter is a combination of RR Lyrae \textit{c} and \textit{d} stars. The same was done for eclipsing binaries with \textit{Contact} (contact and ellipsoidals) and \textit{Non-Contact} stars. The total size and class distribution of the OGLE data-set is shown in table \ref{Tab:classOGLE}. 

\begin{table}
\scriptsize
\caption{OGLE data-set.}\label{Tab:classOGLE}
\begin{tabular}{l p{2.8cm} l}
\hline
\centering
Star Type &Star Class & Size\\
\hline
\multirow{2}{*}{Cepheids} & Ceph & 1,765 \\[0.25ex]
 & Ceph$\_$F & 2,029\\[0.30ex]
$\delta$ Sct & - & 2,444 \\[0.30ex]
\multirow{2}{*}{Eclipsing binaries (EB)} & Contact EB & 883\\[0.25ex]
 & Non-Contact EB & 22,911\\[0.30ex]
\multirow{3}{*}{Long Period Variables (LPVs)} & Mira stars & 572 \\[0.25ex]
 & OSARG & 30,376\\[0.25ex]
 & (SRV) & 4,630 \\[0.25ex]
\multirow{2}{*}{RR Lyrae} & (\textit{ab}) &7,086 \\[0.25ex]
 & \textit{c/d} & 17,146 \\[0.5ex]
\hline
 & \textbf{Total} & 89,842 \\[0.7ex]
\hline
\end{tabular}
\end{table}

\subsection{\textit{Kepler} data-set}

The \textit{Kepler} spacecraft was a telescope designed to observe the northern portion of the Milky Way with the objective to detect Earth-like (in size and possible habitable nature) exoplanets. This project spanned over four years, covering a surface area of 115.6 deg$^{2}$ within the Cygnus and Lyra constellations, and data were released on a quarterly basis. Upon completion of the project, 17 Quarters were covered and over 150,000 stars observed. Two cadences were used during observations, 29.424 minutes for the \textit{long cadence}, which was used in this paper, and 1 minute for the \textit{short cadence}. Planet detection was done by studying the orbiting behaviour of stars using time domain analysis \citep{Koch2010}. Knowledge discovered using data from the mission has resulted in the re-purposing of the telescope after the failure of two reaction-wheels. This was known as the K2 mission \citep{Howell2014} and observations were halted in October 2018.

The \textit{Kepler} data-frame was derived from updated versions of light curves (February 2017) of the analysed quarters of \textit{long cadence} observations and were sourced publicly from MAST\footnote{\url{http://archive.stsci.edu/pub/kepler/lightcurves/tarfiles/}}. Pre-search Data Conditioning Simple Aperture Photometry (PDCSAP) fluxes and their associated errors were used \citep{Stumpe2012, Smith2012} as light curves. PDCSAP was chosen over Simple Aperture Photometry (SAP) fluxes as they are already corrected for electronic effects from the detector and extreme values have been removed. 

As the \textit{Kepler} mission does not have a formalised catalogue, a training set was created using a list of stars that have contributed to the transformation of asteroseismology. Solar-like and red giant stars were taken from studies by \citet{Bedding2011}, \citet{Mathur2012}, \citet{Stello2013}, \citet{Metcalfe2014} and \citet{Creevey2017}. Hot variable stars were sourced from \citet{McNamara2012} and \citet{Balona2015}. Although these authors classified stars as Slow Pulsating B (SPB), $\beta$ Cephei, SPB/$\beta$Cephei hybrids or Maia stars, in this instance they were all labelled as variable B stars. To supplement the size of this class of stars, B stars classified by \citet{Balona2015} from campaign 0 of the K2 mission were also added. $\gamma$ Dor and $\delta$ Scuti stars were from \citet{Bradley2015} with revisions from \citet{VanReeth2015} and \citet{Bowman2016}. Stars that were classed as \textit{hybrids} were removed from the training set and were part of the test set. Eclipsing binaries were from a recent \textit{Kepler} catalogue by \citet{Kirk2016}, using the morphology criterion by \citet{Matijevic2012} to categorise them as contact, detached or semi-detached. RR Lyrae stars were sourced from \citet{Nemec2013MetalField}. Similar to B variables, the size of the RR Lyraes set was supplemented with a sample of RR \textit{ab} stars from \citet{Armstrong2016}. As not all stars are variables, stars classified as \textit{Constants} were used from the work of \citet{McNamara2012}. A total of 11 classes, with red giants and binary stars further divided into sub-classes, was created for the \textit{Kepler} data-set and their distribution is shown in table \ref{Tab:classKepler}.

\begin{table}
\scriptsize
\caption{Class distribution of the \textit{Kepler} data for quarters 14 to 17.}\label{Tab:classKepler}
\begin{tabular}{l p{2.6cm} c c}
\hline
Star Type & Class & Size\\[0.5ex]
\hline
B variables & & 32 \\[0.25ex]
\textit{Constants/Non$\_$Variables} & - & 807 \\[0.25ex]
$\delta$ Sct & - & 1125 \\[0.25ex]
\multirow{2}{*}{Eclising binaries} & Detached & 209 \\[0.25ex]
 & Semidetached/Contact & 328 \\[0.25ex]
$\gamma$ Dor & - & 133 \\[0.25ex]
Other & - & 735 \\[0.25ex]
\multirow{2}{*}{Red Giants} & RGB & 1,828 \\[0.25ex]
& Clump & 1,291 \\[0.25ex]
 RR Lyrae & - &59 \\[0.25ex]
 Solar like &- & 104 \\[0.25ex]
\hline
 & \textbf{Total} & 6,651 \\[0.5ex]
\hline
\end{tabular}
\end{table}

\subsection{Training set}\label{sec:trainingset}

The classifier in this paper was constructed by training a model with a training set, its efficiency was evaluated with a validation set and its ability to generalise assessed with a test set. The data from tables \ref{Tab:classOGLE} and \ref{Tab:classKepler} were then partitioned using a proportion 0.6 for training/learning, 0.2 for model assessment and 0.2 for testing. Since the classes in the data-frame were unevenly dispersed, class ratio were taken into consideration during splitting such that each of the above mentioned sets correlates with the original data-frames.

\begin{table}
\caption{Proportions of the training, validation and test sets for the \textit{Kepler} and OGLE data-sets used.}\label{Tab:datafracs}
\begin{tabular}{l l l l l}
\hline
Data & Training & Validation & Test & Optimisation \\[0.6ex]
\hline
\textit{Kepler} & 4,256 & 1,064 & 2,785 & - \\[0.2ex]
OGLE & 48,873 & 14,375 & 17,969 & 8,625 \\[0.7ex] 
\hline
\end{tabular}
\end{table}

To enhance the performance of the classifier, hyper-parameters associated with the gradient boosted tree (see section \ref{sec:optimisation}) were tuned using training sets. In the case of the OGLE data where the training set consisted of more than 50,000 (the sum of the training 48,873 and optimisation 8,625 sets) stars, a representative sample $(\approx 15\%)$ of that, named the \textit{Optimisation set}, was utilised instead. The decision to use an optimisation set was to reduce model complexities in terms of running times. The size of each of the three data categories used in this study is shown in table \ref{Tab:datafracs} with the last column, designated the \textit{Optimisation set}, introduced for the OGLE data-set. 

\section{Period searching techniques}\label{Sec:period}

\subsection{Data pre-processing}

To accommodate instances where light curves were from different sources, magnitudes (or fluxes) were normalised by dividing by their median and centering around zero by subtracting one. Linear trends were removed prior to feature extraction. Detrending was done to ensure that extracted features derived from light curves were not influenced by any underlying trends. Errors were normalised by dividing by the median of the magnitude/flux. Extreme values were removed with sigma-clipping using a $2.5\sigma$ threshold. Instead of imputation with the mean or median, missing values were removed from light curves. In the case of OGLE, light curves with fewer than 150 points were rejected.

The rejection of \textit{Kepler} light curves was based on the amount of flux within the aperture and crowding from neighbouring stars. This was done using the header keywords \verb"FLFRCSAP" and \verb"CROWDSAP" where light curves with \verb"FLFRCSAP" $ < 0.50$ and \verb"CROWDSAP" $ < 0.80$ were rejected. These header parameters were used as indicators for the quality of the light curves at a specific quarter, that is, to evaluate whether the target star observed was contaminated due to crowding from neighbouring stars and that a sufficient amount of light was enclosed in the aperture. Light curves of stars observed in more than one quarter were concatenated and, to reduce running time during feature extraction, every second point was selected.

\subsection{Period searching}\label{sec:period1}

A key indicator for stellar variability is the period at which the brightness of a star varies over time and this is determined from light curves. Light curves from small scale observations require more robust period searching methods. However, this is not the case for recent large scale surveys as they are continuous for longer periods than the traditional seven day cycle. Consequently, computationally expensive spectral analysis period searching methods such as the Lomb Scargle \citep{Lomb1976, Scargle1982} Periodogram (LSP) can be replaced by faster alternatives. A review by \citet{Graham2013} shows that the selection of a period searching method is primarily driven by completeness and speed. This selection however does not completely disregard the efficiency of the LSP analysis for traditional sinusoidal light curves and its ability to extract multiple periods through the prewhitenning process.

Since \textit{Kepler} light curves were selected on the basis of crowding relative to neighbouring stars and the amount of flux enclosed within the aperture, this led to some quarters being rejected. The LSP was then used to extract periods as it can accommodate the irregular spacing of the data points in light curves. A Python function \verb"astrobase.periodbase", which implements the Lomb-Scargle algorithm, was used as a period extractor. For optimal usage of the frequency estimator, the observed brightness was specified either as magnitude or flux. In case of the latter, the estimator converts them to magnitudes prior to estimations. To confirm the best period estimated ($P_1)$, Friedman's supersmoother \citep{friedman1984} was applied with \verb"gatspy.periodic" \citep{Vanderplas2015,Vanderplas2016}. The supersmoother aims to fit a smoothed linear fit to the raw data prior to estimating the period with the LSP.

\section{Classification workflow}\label{Sec:workflow}

\subsection{Model Training}\label{sec:learning}

There exist a significant number of classification methods  in ML, founded on different algorithms. However, ensemble methods tend to outperform the others most of the time. Ensemble methods are typically defined as a collection of weak learners that form the foundation of a model and predictions are based on \textit{majority voting} \citep{Perrone1993, Freund1995} or \textit{aggregation} of the \textit{base} learners. Commonly known methods include Random Forests \citep[RFs,][]{Breiman2001} and Gradient Boosting Machines \citep[GBMs,][]{Freidman2001} in ML approaches, and deep neural networks (DNNs) that may incorporate models such as Convolutional Neural Networks (CNNs) and Recurrent Neural Networks (RNN) where soft (probabilistic) predictions are employed. 

Assuming that the application of an ensemble method generates $k$ \textit{base} models with the same objective function and hyper-parameters \citep{Dietterich2000}, then
\begin{equation}\label{eq:ensemble}
h_f(x) = \sum\limits^{i = k}_{i = 1} w_i h_i(x),
\end{equation}
where $w_i$ is the weight of the $i$th weak model $h_{i}(x)$ learning from a data-frame $x$ in the ensemble $h_{f}(x)$. Equation \eqref{eq:ensemble} is a common representation of homogeneous learning \textit{ensemble} methods such as Random Forests (RFs), a method composed of a number of trees (in the order of hundreds or thousands). There exist heterogeneous methods that behave in a similar manner as homogeneous ones, however, these involve different models being stacked to form one strong model. The established strong model aims to emphasise the strengths of the individual models. As discussed above, ensemble methods have been used to generate star catalogues for some of the large sky surveys by \citet{Armstrong2016} and \citet{Jayasinghe2018}. There also exist Python-based automatic variable stellar classifiers based on RFs, for example, \citet{Kim2016} and \citet{Bhatti2018} (currently in development stage). The preference for \textit{ensemble} methods is due to their architecture, where the final class/ label assignment is based on predictions of the multiple weak models represented by equation \eqref{eq:ensemble} that form the primary structure of the learner. Furthermore, classification may be hard or probabilistic, which accommodates the overlapping of features that are distinct indicators for different classes of stars. Another advantage of RFs is that they aid with the variance-bias trade-off to which most single learners are exposed. Diversification of equation \eqref{eq:ensemble} is through the manner in which weak learners are constructed. For example, randomised sampling of the training set in RFs results in decision trees in the ensemble learning on ``different" training sets.

\begin{figure*}
\centering
\includegraphics[scale=0.72,valign=b]{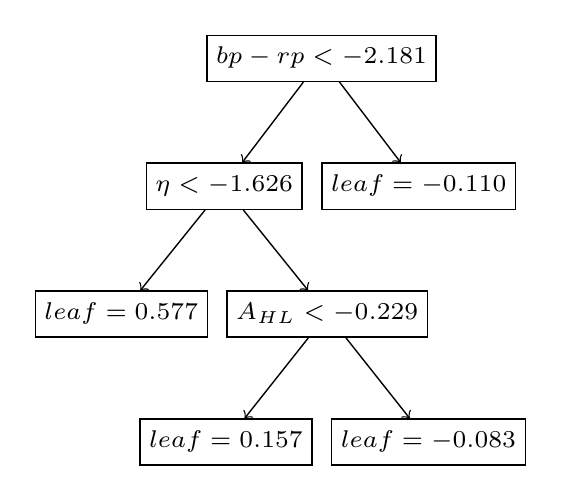}
\includegraphics[scale=0.75,valign=b]{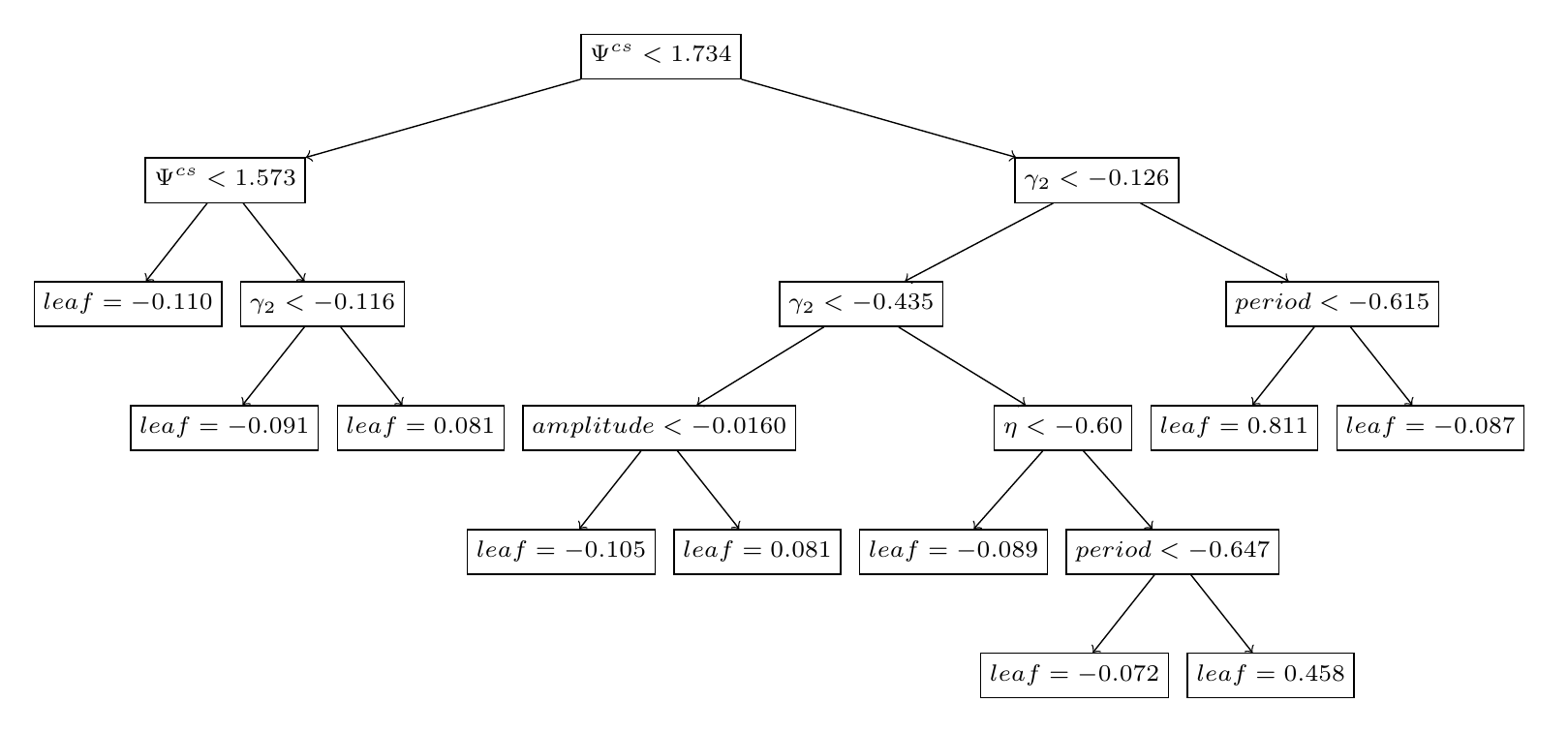}\\
\includegraphics[scale=0.8,height = 6cm,valign=b]{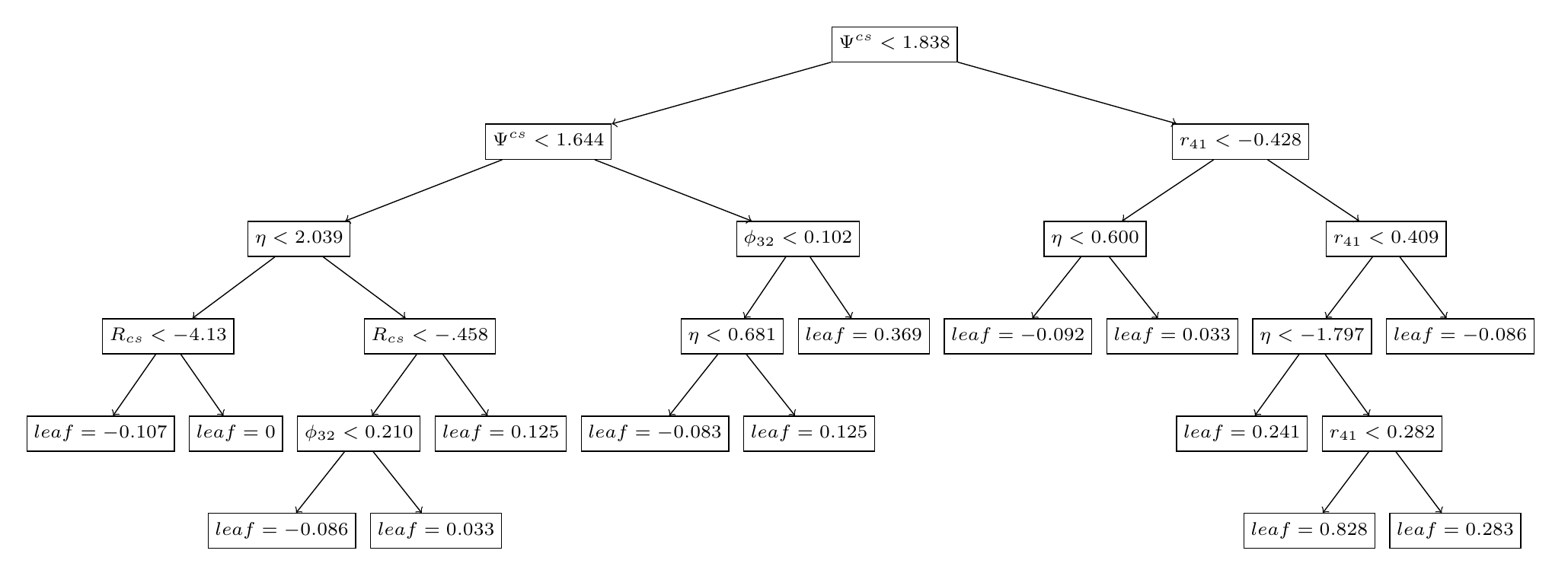}
\caption{Classification trees built with GBMs for the \textit{Kepler} data-set. The top left panel is the first tree of the ensemble, the top right is the $8$th and the third three is represented by the bottom panel. The top nodes are the roots of the tree, internal nodes are those where conditions in the form of features are imposed, resulting in splitting. The nodes where the tree terminates are the leaves and the values in these nodes represent the prediction scores.}\label{Fig:trees}
\end{figure*}

In this paper a resampling method, known as GBMs and called \verb"xgb" in this paper, with similar characteristics to that of RFs, was applied to classify light curve data. GBMs are composed of a number of trees and, unlike RFs, successive trees in the ensemble aim to better the performance of preceding ones. They can be summarised as an adaptive form of training where the $(k + 1)$th learner in the ensemble aims to correct misclassifications from the $kth$ learner. Model improvements are done by iteratively learning from the same training set with modifications applied to the weights of samples. Incorrectly predicted samples are then weighted more compared to true predictions. Consequently, the objective function that the GBMs classifier aims to minimise is,
\begin{equation}\label{eq:gbm}
\mathcal{L}^{(k)} = \sum \limits_{i = 1}^{n}\left[ l(y_i,\hat{y}^{(k - 1)}) + h_k(x) \right] + \Omega(h_k)
\end{equation}
where $n$ is the number of samples in the training set. $l$ is the \textit{loss} function of the classifier and measures the difference between the target $y_i$ and the prediction $\hat{y}_i$. $\Omega(h_k)$ is the regularisation term that tries to reduce the complexity of the model by assigning weights to features to prevent over-fitting. In addition to the behaviour of traditional decision trees, GBMs are able to shrink the amount each tree contributes to the ensemble, thereby allowing successive trees to improve the performance of the model \citep{Chen2016}. Furthermore, sub-sampling of features and samples in the data-frame aid in reducing over-fitting and running times. An example of the evolution of classification trees constructed with GBMs is shown in figure \ref{Fig:trees} where the left panel shows the first tree in the ensemble and the right is the 8th. The mis-classification error of trees improves with the construction of every successive tree. That is, the $8$th will have the least error compared to the first error. 

From figure \ref{Fig:trees}, it is apparent that individual trees in the ensemble are customised. Consequently, trees of GBMs vary slightly from each other in depth and decision boundaries (feature ranking), which is in contrast with RFs where the depth of the trees and feature relevance is the same throughout the ensemble. For example, the splitting of the roots of the trees 1 and 8 in figure \ref{Fig:trees} arise from the $br - rp $ (colour index) and $\Psi^{cs}$ respectively, where $\Psi^{cs}$ is the range of the cumulative sum of a phase-folded light curve \citep{Kim2014}. Decision boundaries used to split the nodes of a tree are then used to measure which features in the data-frame contribute the most during training of a model. One of the commonly used measures is known as information gain and is given by, 
\begin{equation}\label{eq:infogain}
Gain(\textbf{T}, f) = Entropy(\textbf{T}) - \sum \limits_{v \in values(f)} \frac{\mid T_{f = v} \mid}{\mid \textbf{T} \mid} Entropy \left(\textbf{T}_{f = v}\right),
\end{equation}
where $f$ is a feature in the training set $\textbf{T}$, $values(f)$ is a set of possible values of $f$ and $T_{f =v}$ is a subset of features in \textbf{T} with values $v$. The first term of the above equation is the entropy which is the objective function that classification trees aim to maximise. It measures the probability $p_i (y_i = \hat{y_i})$ at which the $\hat{y_i}$ prediction is equal to $y_i$ and is defined as, 
\begin{equation}\label{eq:entropy}
Entropy(\textbf{T}) = - \sum \limits_{i \in values(C)} p_i (y = y_i) \log_2 \left(p_i (y_i = \hat{y_i}) \right),
\end{equation}
where $C$ is a class/target in training set $\textbf{T}$.

Gradient Boosted Machines were implemented with a Python package \verb"xgboostClassifier" (\verb"xgb"), version 0.81 \citep{Chen2016} which is based on \verb"sklearn.base". Due to its ability to scale features and training size (subsampling) during learning \citep{Friedman2002}, \verb'xgb' is rapidly growing as a preferred method for predictive analysis. Furthermore, properties such as feature regularisation may be used for feature selection. Feature regularisation is performed in the form of L1 -fol (\verb"reg_alpha"), an equivalent of the Lasso regularisation \citep{Mairal2012} in regression analysis, and L2 (\verb"lambda"), an equivalent of Ridge regularisation \citep{Warton2008}. These procedures are responsible for penalising features that contribute the least during learning by assigning weights in the interval $[0, 1]$, with minimal contributors assigned weights close to zero (see equation \eqref{eq:gbm}).

As this paper aims to classify at least two classes of stars (multi-class) in a data-frame, the \verb"xgb" classifier was trained such that predictions were done using a probabilistic approach. To classify stars using probabilities, the objective function for multi-class predictions was implemented with \verb"softprob" with final class assignment conditioned using a base score of 0.5 (\ie probabilities greater or equal to 0.5 are considered as a threshold for accepting a candidate into a class). The performance of the classifier during training was monitored with the \textit{loss} function and classification error rate metrics by comparing how the classifier performed when applied to the validation set relative to the training set. The \textit{loss} is the measure of the cost associated with correct prediction and efficient models tend to have a lower cost value, whereas the classification error pays attention to all the objects that were incorrectly classified.

\subsection{Hyper-parameter tuning}\label{sec:optimisation}

Even though the \verb"xgb" classifier tends to perform relatively well with default parameters, if the right hyper-parameters are not used, it may lead to a model with shorter ( or longer) training times which may not be robust. In order to prevent over-fitting and to improve the robustness of the classifier, the parameters listed in table \ref{Tab:opti} were optimised using a shuffled $5-fold$ stratified cross-validation grid search. Detailed descriptions of the parameters in the table are available from the \verb"XGBoost" documentation. Cross-validation was executed during feature selection and parameter tuning was used to account for the nature (imbalanced classes) of the data. Stratification was done such that the \verb"xgb" classifier takes class proportion into consideration during training. Justification of model optimisation was done by comparing the training score of the base model (using parameters in column two of table \ref{Tab:opti}) with that of the optimised model.

\begin{table}
\caption{Extreme boosted optimised hyper-parameters for the \textit{Kepler} and OGLE data-frames.}\label{Tab:opti} 
\begin{tabular}{l c c c c c} 
\hline
Parameter & Default & \multicolumn{3}{c}{Optimised Parameters}\\[0.5ex]
& & \textit{Kepler} & OGLE \\[0.6ex]
\hline
\verb"colsample_bytree" & 1 & 0.6 & 0.7 \\[0.25ex]
\verb"gamma" & 0 & 0.0 & 0.1 \\[0.25ex]
\verb"learning_rate" & 0.2 & 0.2 & 0.1 \\[0.25ex]
\verb"max_depth" & 3 & 12 & 8\\[0.25ex]
\verb"min_child_weight" & 1 & 1 & 1 \\[0.25ex]
\verb"n_estimators" & 100 & 600 & 500\\[0.25ex]
\verb"reg_alpha" & 0 & 0.01 & \verb"1e-5" \\[0.5ex]
\verb"subsample" & 1 & 0.6 & 0.6 \\[0.5ex]
\hline 
\end{tabular}
\end{table}

The base model was trained with a \verb"learning_rate" of 0.2 and the remainder of the parameters in table \ref{Tab:opti} were default values from the \verb"XGBoost" package. In order to find optimal parameters, the model was tested with an array of values for each parameter in table \ref{Tab:opti}. $5-fold$ shuffled cross-validation \verb"GridSearchCV" was used to find the best possible values of hyper-parameters to find an optimal model. A total of eight hyper-parameters were tuned for the \verb"learning_rate" with \verb"n_estimators" being the first to be tuned. Parameters tuned to optimise the \verb"xgb" classifier were, 
\begin{itemize}
\item \verb"colsample_bytree:" The fraction of features used for training per tree. 
\item \verb"gamma:" The minimum that the \textit{loss} function must be for a tree node to be split.
\item \verb"learning_rate:" How much the weights of the classifier must be adjusted with respect to the gradient descent.
\item \verb"max_depth:" The maximum depth that a tree in the ensemble can have. This may vary from tree to tree.
\item \verb"min_child_weight:" The minimum sum of weights of all the samples required to make a further split.
\item \verb"n_estimators:" Number of trees in the ensemble.
\item \verb"reg_alpha:" L1 regularisation on the features in the data-frame.
\item \verb"subsample:" The fraction of the training set used to train a tree. 
\end{itemize}

The \verb"learning_rate" aids in finding the local minimum of the gradient descent such that convergence is reached in a timely manner. The classifier was applied to a number of features that were generated using approaches described in section \ref{sec:Features} below. To reduce model complexities, only a subset of these features was used. Subset selection was employed with Recursive Feature Elimination (RFE), with Correlation Based Feature Selection acting as a supplement. The feature selection strategy, discussed in section \ref{sec:Features}, was applied to respective data-frames and the finalised classifier was trained using the selected subset. Furthermore, an additional step to prevent over-fitting was implemented by introducing an early stopping of 50 to determine the tree limit during predictions. By doing so, the optimal number of trees to be used during predictions on a test set was also determined. 

\subsection{Feature generation}\label{sec:Features}

Since stellar variability is dependent on the behaviour of the brightness as a function of time, light curve features were derived using a combination of domain-based feature extractors. Extracted features can be categorised into periodic and non-periodic features. The former describe the progression of the light curve with time and Fourier decomposition is commonly used in stellar time series analysis. This is a model that fits a sum of sinusoidals to the light curve using the three strongest estimated periods $P_i$, determined as described in section \ref{sec:period1}. The fourth order Fourier model used to estimate periodic features is,
\begin{equation}\label{eq:fourier}
m(t) = m_0(t) + \sum \limits_{i=1}^{3} \sum \limits_{j=1}^{4} \left[ a_{ij} \sin\left( 2 \pi f_{i}j t \right) + b_{ij} \cos\left( 2 \pi f_{i}j t \right)\right],
\end{equation}
where $m(t)$ is the estimated flux/magnitude from the fit and $m_0(t)$, the intercept of the model, is equivalent to the average of the observed brightness. Due to light curve normalisation, $m_0(t) \approx 0$. $f_i = 1/P_i$ (i = 1, 2, 3) is the $ith $ significant frequency. Fitting of the model yields the coefficients $a_{ij}$ and $b_{ij}$,which can be used to calculate the amplitude $A_{ij} = \sqrt{a_{ij} + b_{ij}})$ and phase $ \phi_{ij} = \arctan \left( b_{ij} / a_{ij}\right)$ of the fitted sinusoidal. As the phases $\phi_{ij}$ are not invariant under time translation, amplitude ratios 
\begin{equation}\label{eq:ampratio}
 R_{ji} = \frac{A_{ji}}{A_{i1}},
\end{equation}
and relative phase differences 
\begin{equation}\label{eq:relphase}
 \phi'_{ji} = \phi_{ji} - j \phi_{j1}, 
\end{equation}
are used. Equation \eqref{eq:ampratio} evaluates the changes in amplitude over time whereas equation \eqref{eq:relphase} determines if there is a horizontal shift of the wave-form. Equation \eqref{eq:relphase} assumes that all three periodic variations of the light curves are initially in phase. These periodic features were extracted with the \verb"periodicfeatures" routine nested in the \verb"astrobase.varclass" class. 

Non-periodic features describe the distribution of the fluxes/magnitudes. These include traditional descriptive statistical parameters such as the \verb"meanvariance", \verb"kurtosis" $(\gamma_1)$, \verb"skewness" $(\gamma_2$) and percentile ratios of the brightness. These features were calculated using \verb"astrobase's" \verb"varfeatures.nonperiodic_lightcurve_features" and \verb"upsilon's" \verb"get_features_" functions in Python.

\subsubsection{Data-frame pre-processing}

A total of 29 features that are commonly used as variability indicators were derived from light curves. Data-frames were standardised using, 
\begin{equation}\label{eq:standard}
 X_i' = \frac{X_i - \mu_i}{\sigma_i}
\end{equation}
where $X_i$ is the $ith$ feature in the data-frame, $\mu_i$ and $\sigma_i$ are its average and standard deviation respectively. 

The role of features in ML studies is very crucial as it affects the functionality of constructed algorithms. A review by \citet{Donalek2013} on feature selection with regards to variable stellar properties shows that there exist a number of techniques that can be practised to select a subset of features from $n-$dimensional data-frames. These range from simple methods such as Exploratory Data Analysis (EDA) applied concurrently with correlation coefficients, to complex methods such as Principal Component Analysis (PCA) and wrapping of algorithms. This section will discuss a wrapper method to select the features  that were applied in this study and to assess their importance with respect to the different classes in the data-frame.

Although relatively faster techniques, such as Pearson's correlation coefficients applied as a preliminary selector, could have been applied to select features, they are somewhat independent of the classifier and may not perform well in data sets that contain samples with overlapping properties like variable stars. To ensure that a subset of selected features is plausible, features generated as described in section \ref{sec:Features} were selected by wrapping the optimised model from section \ref{sec:learning} with an algorithm known as recursive feature elimination with cross-validation (RFECV) \citep{Granitto2006}. This is a backward feature selector where the model is recursively trained and features that contribute the least to the model are removed. \verb"sklearn.feature_selection" was used to implement the RFECV tool in Python. $5 - fold$ cross-validation was applied with one feature removed at a time (\verb"step = 1") and accuracy was the metric used to evaluate the feature selector. This was done to reduce complexities associated with feature redundancy during training. Table \ref{Tab:features} shows a finalised list of features used for classification. A detailed description of these features as applied to stellar variability can be found in papers by \citet{Richards2011}, \citet{Kim2014}, \citet{Nun2015} and references therein.

\begin{table}
\caption{Features used to build the \textbf{xgb} classifier for the OGLE and \textit{Kepler} data-frames. Features are primarily from \citet{Richards2011} apart from the $G_{bp - rp}$ and $V I-$ colour indices which are from \citet{Jayasinghe2018a} and \citet{Kim2016} respectively.}\label{Tab:features} 
\begin{tabular}{p{1.75cm} p{2.5cm} p{3.1cm}} 
\hline 
Category &Feature & Data set \\[0.7ex]
\hline
\multirow{2}{*}{Colour index } & $G_{bp - rp}$ & \textit{Kepler} \\[0.25ex]
 & $ V - I $ & OGLE \\[0.25ex]
\addlinespace
\hline
\multirow{3}{*}{Periodic} & \verb"Period" & \textit{Kepler} and OGLE \\[0.25ex]
 & $\phi_{21}$ & OGLE \\[0.25ex]
 & $\phi_{31}$ & OGLE \\[0.25ex]
 & $\phi_{32}$ & \textit{Kepler} \\[0.25ex]
 & $R_{41}$ & \textit{Kepler} and OGLE \\[0.25ex]
 & $R_{43}$ & OGLE \\[0.25ex]
\addlinespace
\hline
\multirow{7}{*}{Non-periodic} & amplitude & \textit{Kepler} \\[0.25ex]
& $A_{HL}$ &\textit{Kepler} \\[0.25ex]
& \verb"Beyond1Std" & \textit{Kepler} \\[0.25ex]
 & $\eta$ & \textit{Kepler} \\[0.25ex]
 & $\gamma_2$ (\verb"kurtosis") &\textit{Kepler} and OGLE \\[0.25ex]
 & \verb"mad" & OGLE \\[0.25ex]
& $\Psi^{cs}$ & \textit{Kepler} and OGLE \\[0.25ex]

& \verb"mag_percentile" & OGLE \\[0.25ex]
& $R{cs}$ & \textit{Kepler} and OGLE \\[0.25ex]
& $\Psi^{\eta}$ & OGLE \\[0.25ex]
& \verb"Shapiro" & OGLE \\[0.7ex]
\hline
\end{tabular}
\end{table}

RFECV was chosen as it selects features based on how they are mapped relative to the class. Although the selected subset may be considered the best, it may contain features that are correlated with each other. Inclusion of highly correlated features may not be beneficial to the model as they have the same effect during learning. Including all of them will only complicate the constructed classifier. Therefore, highly correlated features were removed from the original data-frame. This technique is known as Correlation Based Feature Selection (CBFS) and was done by calculating the correlation coefficient for each of the features and those that returned values greater than 0.65 were removed. The choice for coefficient threshold was based on the fact that this was a supplementary step that aimed to improve on the already selected features. 

To validate the hyper-parameters that were initially determined in section \ref{sec:learning}, the classifier was optimised using features that were selected by RFECV. A finalised classifier was constructed using a subset of features from RFECV and parameters listed in tables \ref{Tab:features} and \ref{Tab:opti} respectively. 

Whilst the strategy in section \ref{sec:Features} may result in a subset of features with a higher validation score, this does not mean that each of the selected features contributes uniformly to the model. If the classifier constructed was simple to interpret, like Support Vector Machines (SVMs), one would be able to extract its coefficients to measure the influence of individual features in the data-frame. However that is not the case for ensemble methods. As discussed in section \ref{sec:learning}, tree-based methods are evaluated using the relevance of features during learning. In \verb'xgb', this is accessed with \verb"model.feature_importance_", which yields the relative importance of each feature in the analysed data-frame. There exists a number of approaches to determine the feature importance and the \verb"gain" is one of the commonly used measures;  evaluating the average loss reduction when a specific feature is used to split a tree. There exist two other measures within the \verb"xgb" package that can be used to measure feature attribution. These are (1) \verb"cover:" this measures the number of times a feature is used to split the data within trees, weighted by the number of training data points that go through splits and (2) \verb"weight:" this is a default in \verb"xgb" and measures the number of times a feature is used to split the data across a tree. These methods return plausible results, but they are inconsistent when working with multi class data-frames as they measure the global feature contribution instead of that for each sample in the training/validation data-frame. An individualistic approach to feature attribution highlights which features are important for respective classes/labels in the data-frame.

Incorporating different measures has been shown to be beneficial for selecting important features. One of the recently developed methods is known as unification. A recently developed unification method known as SHapely Additive exPlanation (SHAP) \citep{Lundberg2017a}, utilises a combination of measures instead of a single measure which is typical of RFs and \verb"XGBoost" algorithms. SHAP analysis was applied as a Python package \verb"shap" \citep{Lundberg2017} to the samples of the validation or test set to return the feature relevance for all the classes. SHAP calculates the influence of features for each of the samples in the validation. The absolute average of the \verb'shap' value was then used to determine the significance of the features with respect to the model constructed. 

\subsection{ Classifier Evaluation}\label{sec:evaluation}

Although a high prediction accuracy in classifiers is desired, this only addresses the \textit{effectiveness} of a classifier (which evaluates generalised capabilities), not its efficiency. To ascertain the latter, the precision-recall values (PRC) \citep{Boyd2013} for each class in the stellar data were calculated. This method of evaluation was used instead of the commonly used area under the receiver operating characteristics (ROC) curves (\verb"roc_area") \citep{Fawcett2006} because it can handle imbalanced data-frames \citep{Saito2015}. Precision-recall values are calculated using the values from a confusion matrix based on predictions (see figures \ref{Fig:CMKepler} and \ref{Fig:OGLECM}). Parameters used to calculate PRC are,
\begin{itemize}
 \item TP: the number of true positives.
 \item TN: the number of true negatives.
 \item FP: the number of false positives.
 \item FN: the number of false negatives
\end{itemize}

Precision is then defined by
\begin{equation}\label{eq:precision}
p = \dfrac{TP}{TP \quad + \quad FP},
\end{equation} 
and recall by,
\begin{equation}\label{eq:recall}
r = \dfrac{TP}{TP \quad + \quad FN}.
\end{equation} 
The accuracy of the positive predictions, expressed as the \textit{f score}
\begin{equation}\label{eq:fscore}
f = 2 \cdot \dfrac{p \times r}{p \quad + \quad r}
\end{equation}
for each class, were then calculated. Equation \eqref{eq:fscore} summarises the positive predictions per class and is then averaged for the classifier. Similar to \verb"roc_auc", \textit{f score} values greater than 0.7 represent an efficient model. Classifier efficiency can be represented by a classification report which shows the precision, recall and \textit{f score} for each of the classes in the analysed data-frame. Classification reports for the \textit{Kepler} and OGLE data are shown in tables \ref{Tab:classreportKepler} and \ref{Tab:classreport} respectively.  The OGLE report showed a weighted \textit{f score} average of 0.97, with contact eclipsing binaries having the lowest \textit{f score} value (0.50) and RR Lyrae \textit{ab} (\textit{f score} = 0.99) having the highest value. In addition to calculating the $f score$ through classes, the average $f score$ of the \verb"xgb" classifier was determined by wrapping it with a binary classifier \verb"OnevsRestClassifier". To prevent information leakage, techniques in this section and section \ref{sec:Features} were applied onto the validation sets discussed in section \ref{sec:trainingset}.%

\section{Results}\label{Sec:results}

\subsection{Classifier Performance}

\begin{figure}
\includegraphics[width = 8.5 cm, height = 12.0cm]{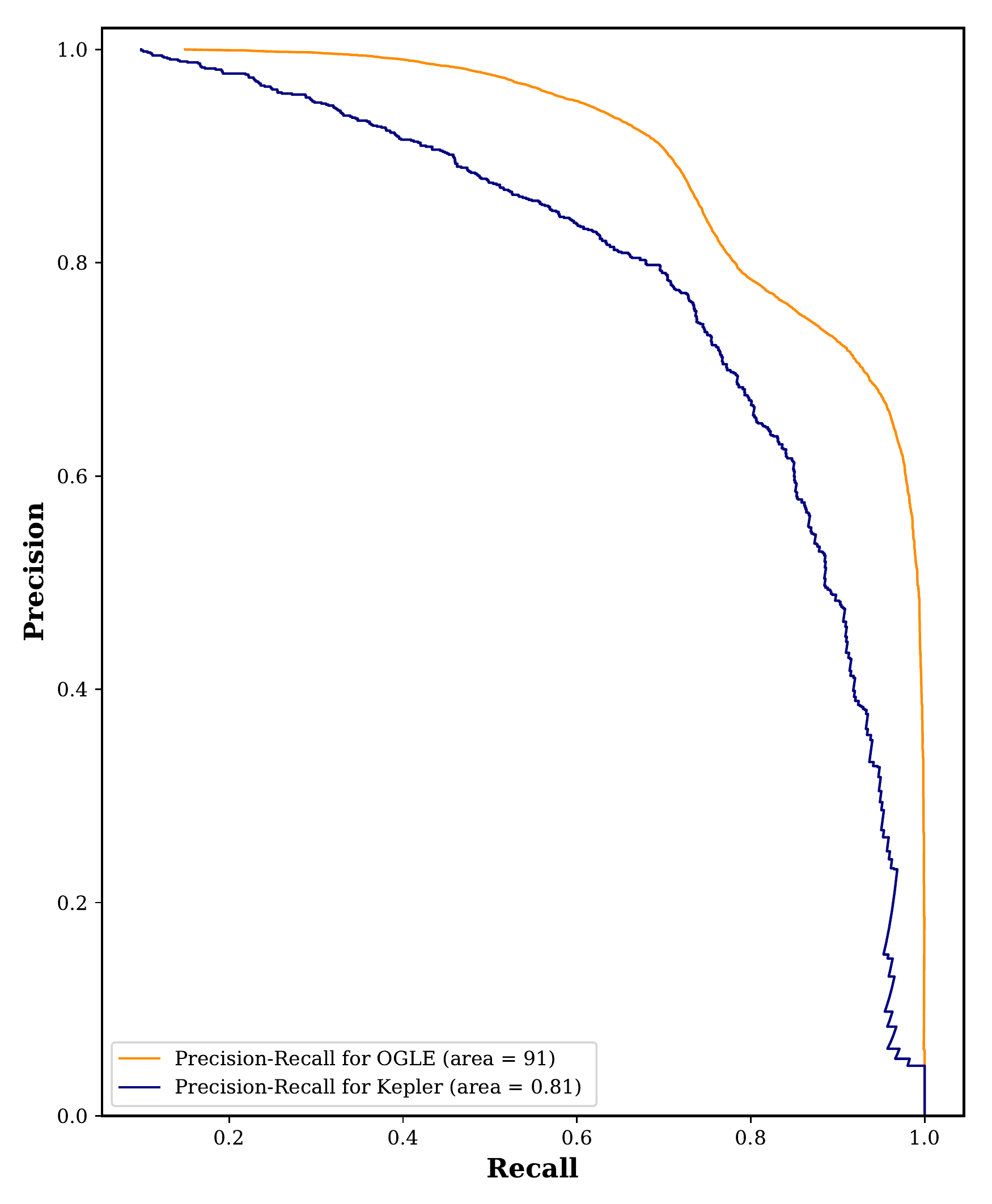}
\caption{Precision as a function of recall curves for the classifier when applied to the two data-sets in section \ref{Sec:Data}. The results were obtained by testing the prediction on the validation sets. The average precision values indicate an efficient model.}\label{Fig:PRcurve}
\end{figure}

Although the proposed classification workflow discussed in the previous section has proven computationally expensive with intermediate steps performed with cross-validation, the \verb"xgb" classifier constructed with parameters in table \ref{Tab:opti} is efficient and effective. The feature subset selection resulted in significantly fewer features than the original size of the data-frames. The SHAP analysis for the OGLE data (see figure \ref{Fig:OGLEFR}) shows a clear distinction between the most relevant features and the least contributors. A similar distinction was observed for the \textit{Kepler} data which suggested that Fourier parameters are the least contributors for model. Furthermore, the classifier had an average precision of 0.86, with the OGLE data having the higher (0.91) and the \textit{Kepler} data the lower (0.81) values (see figure \ref{Fig:PRcurve}). Precision values for respective data show a fraction of correctly classified stars relative to the sum of correctly and incorrectly labelled stars. These values were consistent with the $f scores$ of the classifier from the test set classification reports of the analysed data. In both instances, restricting the number of trees during predictions instead of using the entire ensemble proved sufficient. An average of $6\%$ of the trees in the ensemble were used to predict the test sets.

Using the parameters in table \ref{Tab:opti} resulted in training scores of at least $90 \%$ in both data-frames. This clearly shows that the constructed \verb"xgb" classifier is effective and efficient. Consequently, it was used to classify data from the surveys discussed in section \ref{Sec:Data}. 

We wanted to assess whether the classifier could be successfully applied to a data-frame created by combining different sources. To this end, we sampled stars common to the \textit{Kepler} and OGLE data-frames to produce a \textit{new} data-frame consisting of periodic and non-periodic features only. To homogenise the data-frame, it was normalised such that each feature in the data-frame had a range of one. Whilst this was a competent classifier, which resulted in a weighted $f score$ of 0.79, it had a relatively poor generalisation on cepheids stars. The number of trees used in this sample was similar to those of the \textit{Kepler} data with an optimal depth of eight. The ability to classify each of the classes from this \textit{multi}-source data-frame is shown in table \ref{Tab:classreporthybrid}. Evaluation of the predictions of the training set relative to the validation set suggested that predictions could be obtained with one tree. However, to ensure a plausible prediction, an additional five trees were added. In this instance, the SHAP analysis showed that flux/magnitudes percentiles (\verb"mag_percentiles") and $\eta$ were the top contributors to the model and the oscillatory period was the only periodic feature selected in the finalised sub-set. 

\begin{table}
\scriptsize
\caption{Classification report for the eight classes in the data-frame sampled from the \textit{Kepler} and OGLE data-frame. Training was done using 7,012 stars and tested on 1,753 stars.}\label{Tab:classreporthybrid} 
\begin{tabular}{l l c c } 
\hline
Type & Precision & Recall & $ f \quad score$ \\[0.7ex]
\hline
Cepheids & 0.00 & 0.00 & 0.00 \\[0.25ex]
 $\delta$ Sct & 0.90 & 0.76 & 0.82 \\[0.25ex]
Eclipsing Binaries & 0.75 & 0.73 & 0.74 \\[0.25ex]
$\gamma$ Dor & 0.33 & 0.22 & 0.27 \\[0.25ex]
LPVs & 0.84 & 0.99 & 0.91\\[0.25ex]
\textit{Non Variables} & 0.95 & 0.77 & 0.85\\[0.25ex]
RR Lyrae & 0.79 & 0.90 & 0.84\\[0.25ex]
Solar like & 0.29 & 0.38 & 0.33\\[0.25ex]
 \hline
\textbf{weighted avg} & 0.79 & 0.81 & 0.79 \\[0.7ex]
\hline 
\end{tabular}
\end{table}
Results from these data-sets suggests that, in addition to the above workflow, the size and nature of the training set used to construct a classification model drives its performance. This implies models built from larger, homogeneous data may be more desirable. However, homogeneity might result in classifiers that are insufficiently diverse, producing model bias. This implies that a complete re-training might be a necessary prior to classifying a new data-set. Such an outcome would be far from ideal, in view of the plethora of high-dimensional data expected in the future. 
\subsection{Catalogue predictions}

\subsubsection{\textit{Kepler} data}

\begin{figure*}
\includegraphics[scale = 0.6]{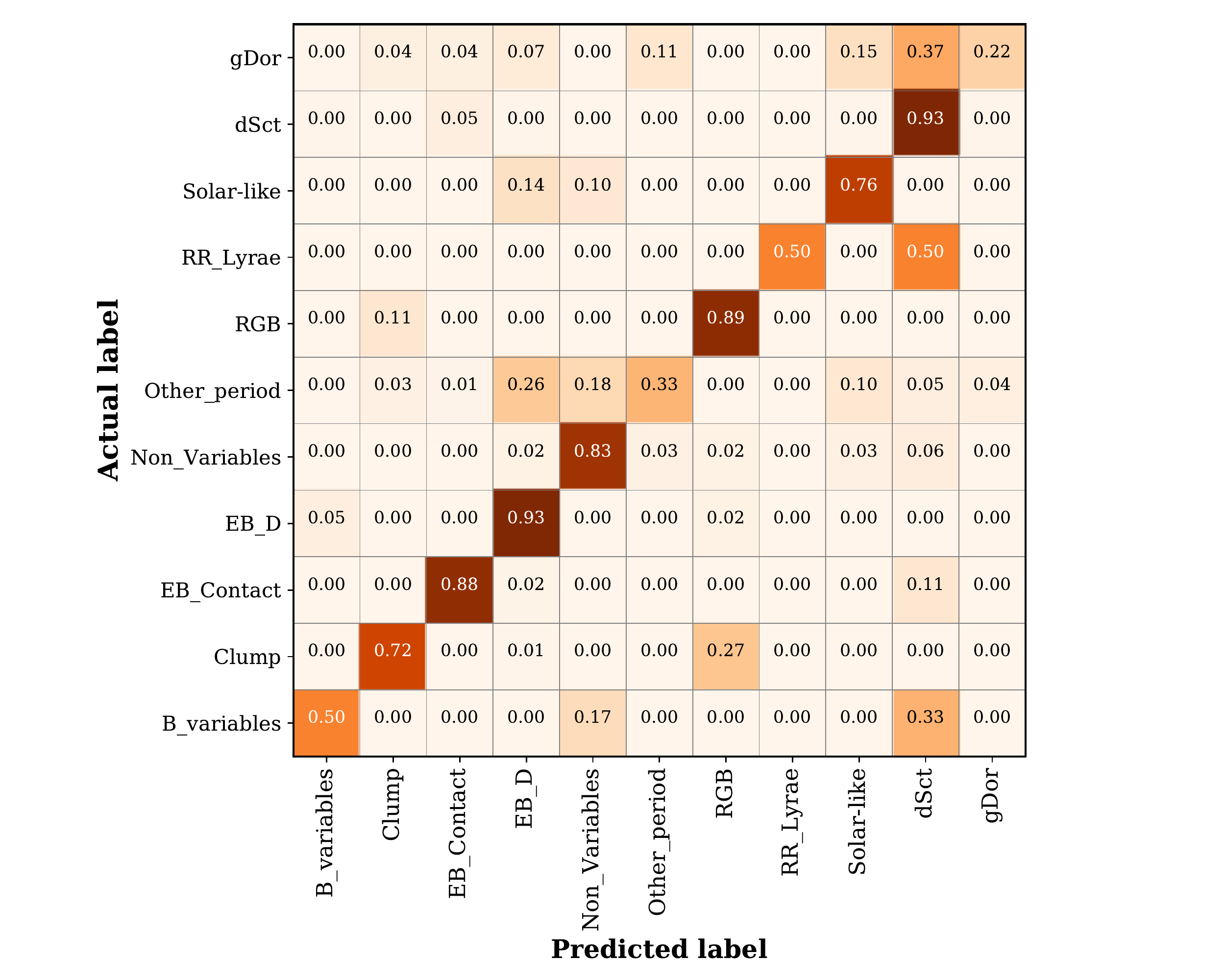}
\caption{Confusion matrix for the classes in the \textit{Kepler} data. The diagonal squares show positively predicted samples with values greater than 0.5 showing a higher recall \eqref{eq:recall}.}\label{Fig:CMKepler}%
\end{figure*}

As discussed above, the classifier built with the \textit{Kepler} data-frame was competent and resulted in an average precision of 0.8 from the validation set and weighted $f score$ of 0.77 from the 11 classes of stars in the data-frame. The results were obtained from training the \verb"xgb" on the data-frame consisting of 11 features and 4,256 stars. The competency of the \verb"xgb" classifier was demonstrated when it was applied to the slightly diverse \textit{Kepler} data. Implementation of the classifier onto this data-frame resulted in an accuracy of $83 \%$. From figure \ref{Fig:CMKepler} and \textit{ f score} values in table \ref{Tab:classreportKepler}, it is clear that seven of the 11 labels in the data-frame were comparatively easy for the classifier to learn from. Half of RR Lyraes and $33\%$ of the B variables were misclassified as $\delta$ Sct stars. The classifier performed poorly with respect to $\gamma$ Dor stars, with an accuracy of $22\%$. Classes with accuracies above $75\%$ had a higher proportion of stars to learn from during training of the model and the three poorly-classified classes all suffered from small sample sizes. This was due to the limited number of labelled stars of these types observed in the selected quarters. To overcome this, the introduction of more features in the data-frame, or extension of the training set search to include more \textit{Kepler} quarters, may aid in improving the efficiency of the \verb"xgb" classifier in this data-frame.

\begin{table}
\scriptsize
\caption{Classification report for the 11 classes in the \textit{Kepler} data-set. The confusion matrix for this report is shown in figure \ref{Fig:CMKepler}}\label{Tab:classreportKepler} 
\begin{tabular}{l l c c c } 
\hline
Type & Class & Precision & Recall & $ f \quad score$ \\[0.7ex]
\hline
B variables &- & 0.60 & 0.50 & 0.55 \\[0.25ex]
 $\delta$ Sct & - &0.84 & 0.94 & 0.88 \\[0.25ex]
 \multirow{2}{*}{Eclipsing Binaries} & Detached &0.46 & 0.93 & 0.62 \\[0.25ex]
 & SD/Contact & 0.81 & 0.89 & 0.85 \\[0.25ex]
$\gamma$ Dor & - &0.54 & 0.15 & 0.24 \\[0.25ex]
\textit{Non Variables} & - & 0.79 & 0.91 & 0.84\\[0.25ex]
\textit{Other periodics} & - & 0.86 & 0.44 & 0.58\\[0.25ex]
\multirow{2}{*}{Red giants} & RGB & 0.79 & 0.90 & 0.84 \\[0.25ex]
 & Clump & 0.81 & 0.66 & 0.73\\ [0.25ex]
RR Lyrae &- & 1.00 & 0.75 & 0.86\\[0.25ex]
Solar like &- &0.77 & 0.48 & 0.59\\[0.25ex]
\hline
& \textbf{weighted avg} & 0.80 & 0.79 & 0.77 \\[0.7ex]
\hline 
\end{tabular}
\end{table}
The colour index (in this instance GAIA $bp - rp$) and $\eta$ were the strongest contributing feature in the classifier for this data-frame, where $\eta$ measures the degree of monotonic increase or decrease of flux/magnitude in a long-term baseline \citep{Kim2014}. The results obtained and size of the training set suggests that the inclusion of other astrophysical parameters such as mass ($M$) and proxies for temperature and composition (\eg 2MASS colour indices and metallicity) may make classification more efficient, if available. Figure \ref{Fig:CMKepler} and table \ref{Tab:classreportKepler} summarise the results obtained with the \verb"xgb" classifier for the \textit{Kepler} data-frame. The relevance of the features calculated with the SHAP analysis selected the Gaia colour index as the most influential feature and Fourier parameters as the least contributors to the trained model.

\subsubsection{OGLE data}

\begin{figure*}
\centering
\includegraphics[width=15.75cm]{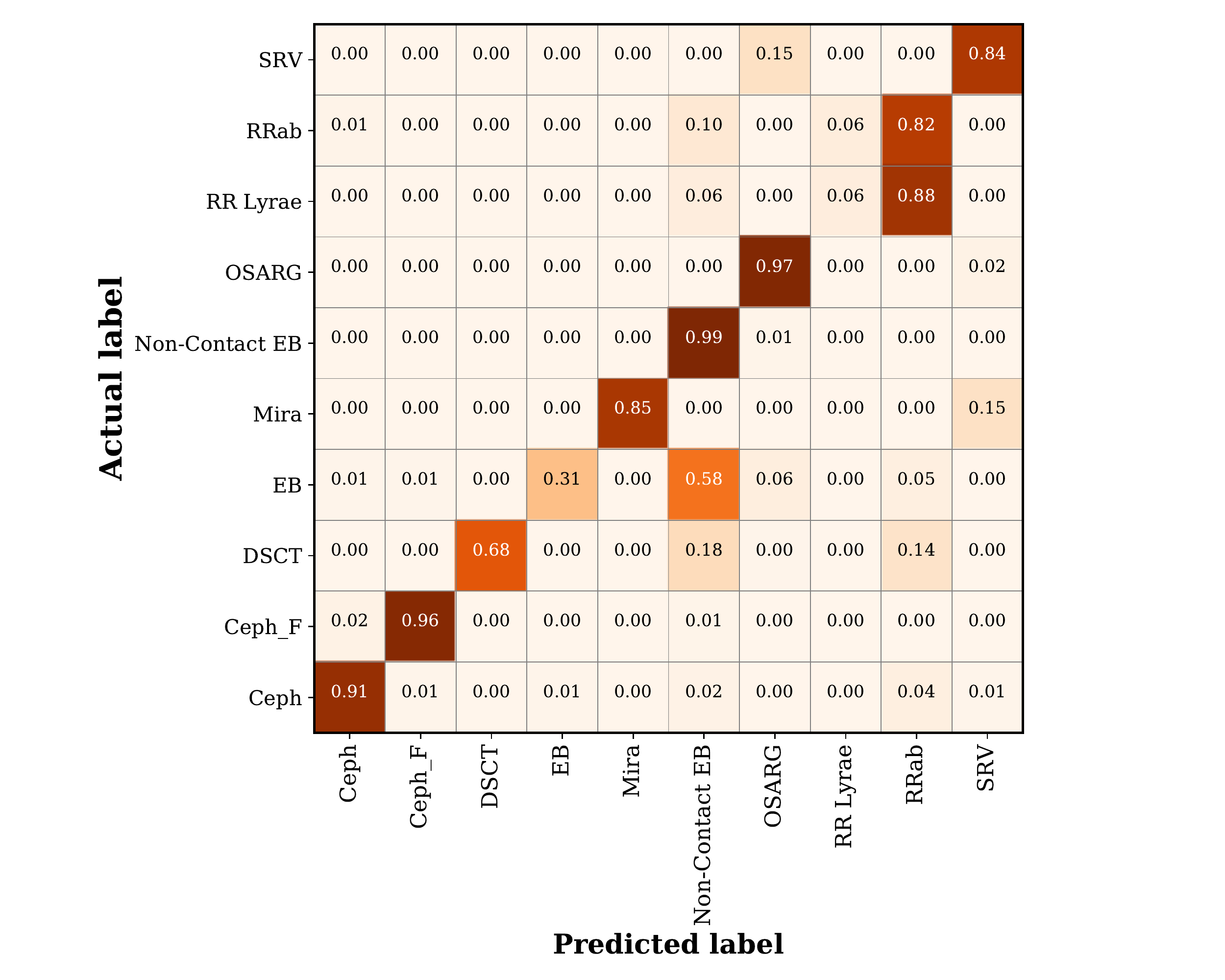}
\caption{Confusion matrix for the OGLE data-set with an accuracy of $96.7 \%$. The diagonal darker shades show correctly classified stars. The numbers are the proportion of the correctly classified stars relative to the class in the test set.}\label{Fig:OGLECM}
\end{figure*}

\begin{figure*}
\centering
\includegraphics[scale=0.58, width = 10.0cm]{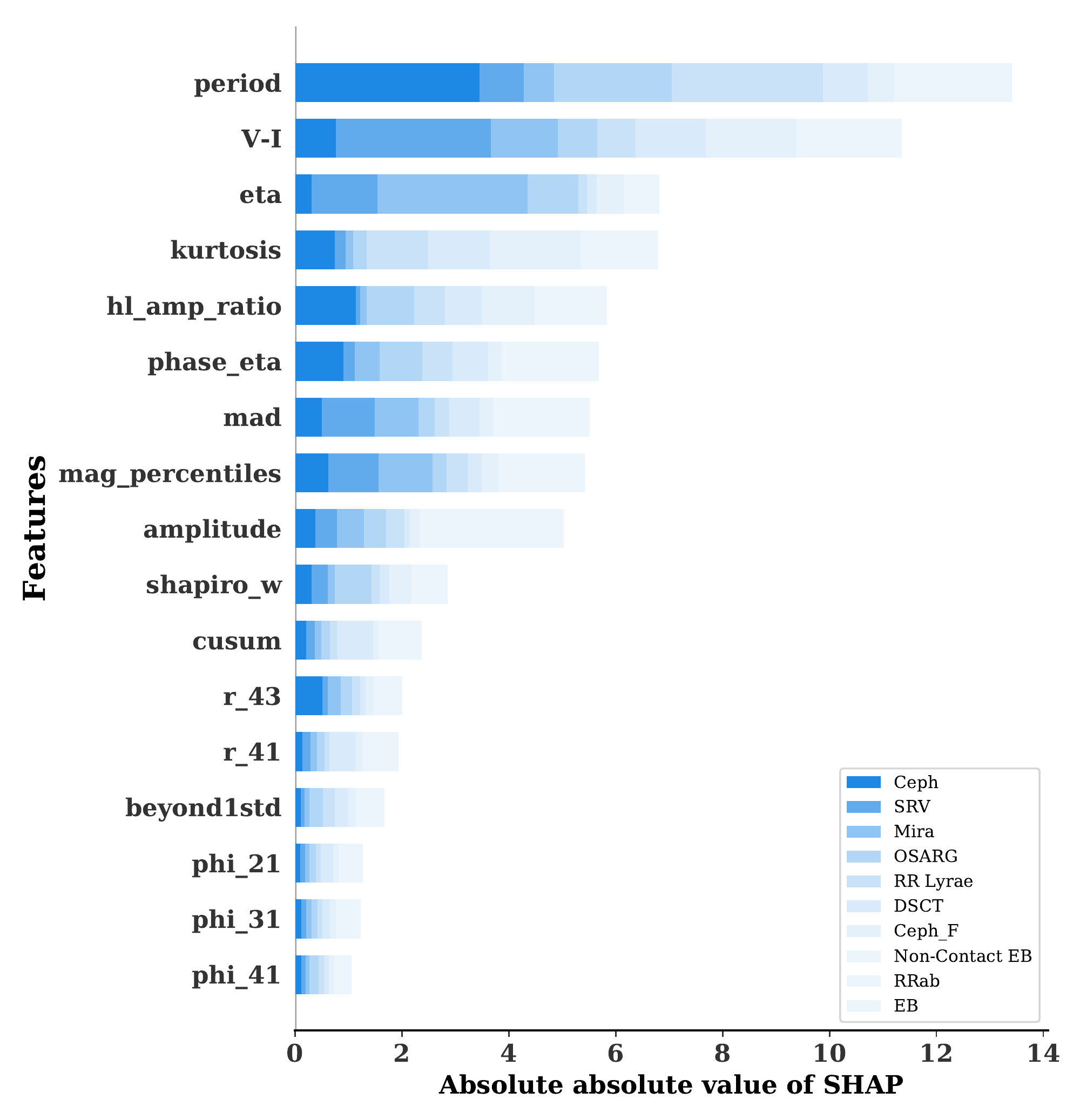}
\caption{Feature importance of the classifier when applied to the OGLE data-frame. The accuracy was $85.2 \%$ and generalisation abilities are shown in figure \ref{Fig:OGLECM}.}\label{Fig:OGLEFR}
\end{figure*}

Generalisation and efficiency of the classifier for the OGLE data-frame were determined by testing it using star types and star classes respectively. In both instances a training score of $\approx 99 \%$ was obtained. From this approach, it is apparent that the number of trees required to make predictions was influenced by the number of labels to which a classifier could assign a particular sample/object. An additional factor was the \textit{f score} value of the smallest number of categories in the data frame. In this instance, a comparison of using star classes relative to types showed an improvement of 0.02 of the $f score$ (which was in favour of classes) was achieved. Considering the minimal difference in the results, star classes were used in the final model. This resulted in a model with the parameters listed in table \ref{Tab:opti}, with predictions of the test set performed with 40 trees. A summary of results obtained when constructing the \verb"xgb" classifier with the OGLE data-frame appear in figures \ref{Fig:OGLECM} and \ref{Fig:OGLEFR} and table \ref{Tab:classreport}. 

The SHAP analysis discussed in section \ref{sec:Features} suggested that, for the OGLE data-frame, traditional features like \verb"period" and $V – I$ colour were key features, while Fourier parameters (equations \eqref{eq:ampratio} and \eqref{eq:relphase}) were only minimal contributors. The same analysis also showed that the top features were pivotal for identifying RR Lyrae \textit{ab} and OSARG LPV stars. This finding is illustrated by the various shades of blue in figure \ref{Fig:OGLEFR}. Despite the fact that the classifier performed poorly with respect to contact or ellipsoidal eclipsing stars, the mis-classification was primarily within the eclipsing type, which implies that this may be overcome by introducing more data to learn from features beneficial to this class.

\begin{figure}
\centering
\includegraphics[width=8.35cm]{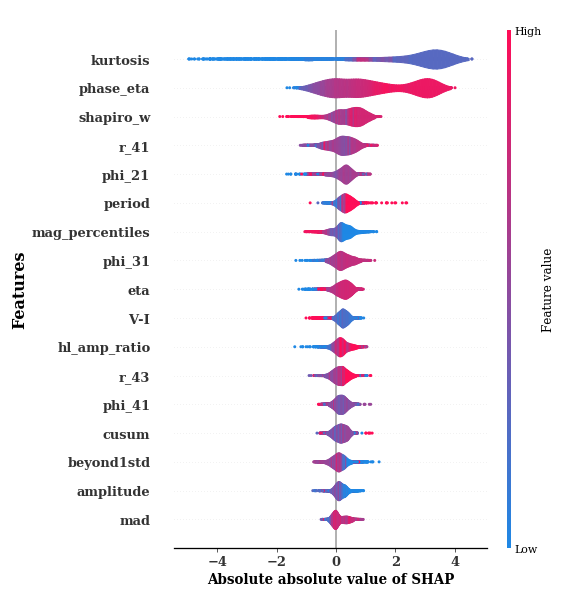}
\caption{Feature relevance of binary classification of eclipsing binary stars in the OGLE data-frame. This violin plot shows $\gamma_2$ was the most relevant feature. }\label{Fig:FSOGLEEB}
\end{figure}

The preference for non-periodic features was highlighted when the classifier was applied to a single type of star which can be further divided into classes. The feature importance for binary classification of eclipsing binaries in the OGLE data-set, as displayed in figure \ref{Fig:FSOGLEEB}, shows that the oscillatory period is the sixth significant feature and the $V - I$ colour index is amongst the least contributors. The descriptive statistics parameter kurtosis $(\gamma_2)$ is the best feature for this classification. Furthermore, the classifier showed a similar effect where it performs poorly with respect to distinguishing contact from non-contact eclipsing variables. Overall, the efficiency of the binary/three-class \verb"xgb" classifier was consistent with that of a 10 class data-frame of the OGLE data.

\begin{table}
\scriptsize
\caption{Classification report for the nine classes in the OGLE data-set when applying the \textbf{xgb} classifier to a 34,842 test set.}\label{Tab:classreport} 
\begin{tabular}{l l c c c } 
Type & Class & Precision & Recall & $ f \quad score$ \\[0.7ex]
\hline
\multirow{2}{*}{Classical Cepheids} & Ceph & 0.84 & 0.91 & 0.88 \\[0.25ex]
 & Ceph$\_$F & 0.91 & 0.96 & 0.96\\[0.25ex]
 $\delta$ Sct & &0.99 & 0.68 & 0.81\\[0.25ex]
 \multirow{2}{*}{Eclipsing Binaries} & EB$\_$C &0.83 & 0.31 & 0.45 \\[0.25ex]
 & EB$\_$NC & 0.96 & 0.85 & 0.90\\[0.25ex]
\multirow{3}{*}{LPVs} & Mira &0.96 & 0.85 & 0.90\\[0.25ex]
 & OSARG & 0.97 & 0.97 & 0.93\\[0.25ex]
 & SRV & 0.82 & 0.84 & 0.83 \\[0.25ex]
\multirow{2}{*}{RR Lyrae} & (\textit{ab}) & 0.68 & 0.82 & 0.74 \\[0.25ex]
 & \textit{c} and \textit{d} & 0.29 & 0.06 & 0.10\\ [0.6ex]
 \hline
& \textbf{weighted avg} & 0.97 & 0.97 & 0.97 \\[0.7ex]
\hline 
\end{tabular}
\end{table}

\section{Conclusions}

A resampling ensemble method of machine learning in the form of extreme boosting was applied to a sample of light curves taken from two sources: OGLE and \textit{Kepler}. Features were derived with similar methodologies across both sources. This was done by implementing an iterative classification workflow typical of Data Science. The methods applied can be summarised as a set of weak classification trees iteratively learning from the same training set, with hyper-parameters optimised for each of the data-frames. Key differences between the \verb"xgb" classifier and other tree-like methods are scalability during training and different orientations of the trees in the ensemble. We conclude that the feature categories listed in Table \ref{Tab:features} may require an extension to enable optimal discrimination between labels. Furthermore, we argue that, to reduce the number of features in the data-frame, hybrid feature subset selection such as the one proposed may be applied.

Depending on the size of the training set, a fraction of each set ($\approx 10\%$) was used to optimise the classifier. We introduce this as an \textit{Optimisation set} that aids in dealing with very large training sets. To attest the efficiency of the constructed model, we evaluated it on a validation set $(\approx 20\%$ of the data-frame) using precision recall curves shown in figure \ref{Fig:PRcurve} which confirmed this. The classification work flow discussed in section \ref{Sec:workflow} and the generalisation of the classifier were performed using a data splitting ratio of 60 (Training):20 (Validation):20 (Test).

Hyper-parameters of the trained model were executed through a shuffled $5-fold$ stratified cross-validation grid search. A threshold probability value greater than or equal to 0.5 was used for accepting a candidate star into a particular class. During the process, the optimal number of trees to be used during predictions was also determined. Optimal features to be used for classification were determined by recursive training of the classifiers using backward step-wise selection. 

Subsequent to the training and tuning procedures, the final classifier was applied to two distinct data-sets containing variable stellar data, \ie \textit{Kepler} and OGLE. Given the selection criteria forced on the \textit{Kepler} data-frame, it contained the lowest number of stars in this study. Two independent data-frames derived primarily from 96,493 stellar light curves were analysed in this study. Our results highlight the flexibility of the \verb"xgb" classifier with minimal re-training, suggesting it can be transferred to other unseen data.The efficiency of the classifier was determined using the average precision and recall values by wrapping the classifier with a binary classifier and calculating the \textit{f score} values of each of the classes in the respective data-frame. The accuracy of the positive predictions of each final classifier (expressed as the \textit{f score}) were 0.81 and 0.91 for the \textit{Kepler} and OGLE data respectively.

We conclude that an ensemble machine learning procedure using extreme boosting can achieve very high efficiencies for classifying variable stars in both ground-based and space-based large surveys. We propose that, although some of the intermediate stages such as hybrid feature selection techniques and tree limit for predictions in the classification may be expensive with respect to time and computation, they have proven to be instrumental in producing an efficient model. The results obtained suggest that the \verb"xgb" classifier can be incorporated into highly complex methods such as feature extraction with deep learning, and produce highly accurate results. This further proved that this procedure deserves serious consideration for classification of variable stars in upcoming survey programmes. 

\section*{Acknowledgements}

This study was partly funded by the James Cook University's Graduate Research School Research Training Program and College of Science and Engineering Higher Degree funding. The authors would like to thank the TESS Data for Asteroseismology working group for their contribution through discussions in Leuven in November 2018. 

The research leading to these results has (partially) received funding from the European Research Council (ERC) under the European Union's Horizon 2020 research and innovation programme (grant agreement N$^\circ$670519: MAMSIE), from the KU\,Leuven Research Council (grant C16/18/005: PARADISE), from the Research Foundation Flanders (FWO) under grant agreement G0H5416N (ERC Runner Up Project), as well as from the BELgian federal Science Policy Office (BELSPO) through PRODEX grant PLATO.




\bibliographystyle{mnras}
\bibliography{GBGC} 

\begin{thebibliography}{}
\makeatletter
\relax
\def\mn@urlcharsother{\let\do\@makeother \do\$\do\&\do\#\do\^\do\_\do\%\do\~}
\def\mn@doi{\begingroup\mn@urlcharsother \@ifnextchar [ {\mn@doi@}
  {\mn@doi@[]}}
\def\mn@doi@[#1]#2{\def\@tempa{#1}\ifx\@tempa\@empty \href
  {http://dx.doi.org/#2} {doi:#2}\else \href {http://dx.doi.org/#2} {#1}\fi
  \endgroup}
\def\mn@eprint#1#2{\mn@eprint@#1:#2::\@nil}
\def\mn@eprint@arXiv#1{\href {http://arxiv.org/abs/#1} {{\tt arXiv:#1}}}
\def\mn@eprint@dblp#1{\href {http://dblp.uni-trier.de/rec/bibtex/#1.xml}
  {dblp:#1}}
\def\mn@eprint@#1:#2:#3:#4\@nil{\def\@tempa {#1}\def\@tempb {#2}\def\@tempc
  {#3}\ifx \@tempc \@empty \let \@tempc \@tempb \let \@tempb \@tempa \fi \ifx
  \@tempb \@empty \def\@tempb {arXiv}\fi \@ifundefined
  {mn@eprint@\@tempb}{\@tempb:\@tempc}{\expandafter \expandafter \csname
  mn@eprint@\@tempb\endcsname \expandafter{\@tempc}}}

\bibitem[\protect\citeauthoryear{Armstrong et~al.,}{Armstrong
  et~al.}{2016}]{Armstrong2016}
Armstrong D.~J.,  et~al., 2016, \mn@doi [\mnras] {10.1093/mnras/stv2836}, 456,
  2260

\bibitem[\protect\citeauthoryear{Bailer-Jones, Rybizki, Fouesneau, Mantelet  \&
  Andrae}{Bailer-Jones et~al.}{2018}]{Bailer2018}
Bailer-Jones C. A.~L.,  Rybizki J.,  Fouesneau M.,  Mantelet G.,   Andrae R.,
  2018, \aj, 156, 58

\bibitem[\protect\citeauthoryear{Ball \& Brunner}{Ball \&
  Brunner}{2009}]{Ball2009}
Ball N.~M.,  Brunner R.~J.,  2009, \mn@doi [International Journal of Modern
  Physics D] {10.1142/S0218271810017160}, 19, 1049

\bibitem[\protect\citeauthoryear{Balona et~al.,}{Balona
  et~al.}{2011}]{Balona2011e}
Balona L.~A.,  et~al., 2011, \mn@doi [\mnras]
  {10.1111/j.1365-2966.2011.18311.x}, 413, 2403

\bibitem[\protect\citeauthoryear{Balona, Baran, Daszy{\'{n}}ska-Daszkiewicz  \&
  De~Cat}{Balona et~al.}{2015}]{Balona2015}
Balona L.~A.,  Baran A.~S.,  Daszy{\'{n}}ska-Daszkiewicz J.,   De~Cat P.,
  2015, \mn@doi [\mnras] {10.1093/mnras/stv1017}, 451, 1445

\bibitem[\protect\citeauthoryear{Barban et~al.,}{Barban
  et~al.}{2007}]{Barban2007}
Barban C.,  et~al., 2007, \mn@doi [\aap] {10.1051/0004-6361:20066716}, 468,
  1033

\bibitem[\protect\citeauthoryear{Bedding et~al.,}{Bedding
  et~al.}{2011}]{Bedding2011}
Bedding T.~R.,  et~al., 2011, \nat, 471, 1

\bibitem[\protect\citeauthoryear{Beichman et~al.,}{Beichman
  et~al.}{2014}]{Beichman2014}
Beichman C.,  et~al., 2014, \mn@doi [\pasp] {10.1086/679566}, 126, 1134

\bibitem[\protect\citeauthoryear{Bhatti, Bouma  \& Wallace}{Bhatti
  et~al.}{2018}]{Bhatti2018}
Bhatti W.,  Bouma L.~G.,   Wallace J.,  2018,
  {{\textbackslash}texttt{\{}astrobase{\}}}, \mn@doi{10.5281/zenodo.1185231},
  \url {https://doi.org/10.5281/zenodo.1185231}

\bibitem[\protect\citeauthoryear{Blomme et~al.,}{Blomme
  et~al.}{2011}]{Blomme2011}
Blomme J.,  et~al., 2011, Monthly Notices of the Royal Astronomical Society,
  418, 96

\bibitem[\protect\citeauthoryear{Borne}{Borne}{2009}]{Borne2009}
Borne K.~D.,  2009, eprint, arXiv:0909.3892

\bibitem[\protect\citeauthoryear{Borne}{Borne}{2013}]{Borne2013}
Borne K.~D.,  2013, in , Planets, Stars and Stellar Systems Volume 2:
  Astronomical Techniques, Software, and Data.
Springer, Chapt.~9, pp 403--444

\bibitem[\protect\citeauthoryear{Bowman, Kurtz, Breger, Murphy  \&
  Holdsworth}{Bowman et~al.}{2016}]{Bowman2016}
Bowman D.~M.,  Kurtz D.~W.,  Breger M.,  Murphy S.~J.,   Holdsworth D.~L.,
  2016, \mn@doi [\mnras] {10.1093/mnras/stw1153}, 460, 1970

\bibitem[\protect\citeauthoryear{Boyd, Eng  \& Page}{Boyd
  et~al.}{2013}]{Boyd2013}
Boyd K.,  Eng K.~H.,   Page C.~D.,  2013, in , Vol. 8190 LNAI, Lecture Notes in
  Computer Science (including subseries Lecture Notes in Artificial
  Intelligence and Lecture Notes in Bioinformatics).
Springer Verlag, pp 451--466, \mn@doi{10.1007/978-3-642-40994-3_29}

\bibitem[\protect\citeauthoryear{Bradley, Guzik, Miles, Uytterhoeven,
  Jackiewicz  \& Kinemuchi}{Bradley et~al.}{2015}]{Bradley2015}
Bradley P.~A.,  Guzik J.~A.,  Miles L.~F.,  Uytterhoeven K.,  Jackiewicz J.,
  Kinemuchi K.,  2015, \mn@doi [\aj] {10.1088/0004-6256/149/2/68}, 149

\bibitem[\protect\citeauthoryear{Breiman}{Breiman}{2001}]{Breiman2001}
Breiman L.,  2001, \mn@doi [Machine Learning] {10.1023/A:1010933404324}, 45, 5

\bibitem[\protect\citeauthoryear{Brett, West  \& Wheatley}{Brett
  et~al.}{2004}]{Brett2004}
Brett D.~R.,  West R.~G.,   Wheatley P.~J.,  2004, \mn@doi [\mnras]
  {10.1111/j.1365-2966.2004.08093.x}, 353, 369

\bibitem[\protect\citeauthoryear{Casanellas \& Lopes}{Casanellas \&
  Lopes}{2010}]{Casanellas2010}
Casanellas J.,  Lopes I.,  2010, \mn@doi [\mnras]
  {10.1111/j.1365-2966.2010.17463.x}, 410, 535

\bibitem[\protect\citeauthoryear{Chaplin et~al.,}{Chaplin
  et~al.}{2014}]{Chaplin2014}
Chaplin W.~J.,  et~al., 2014, \mn@doi [\apjs] {10.1088/0067-0049/210/1/1}, 210

\bibitem[\protect\citeauthoryear{Chen \& Guestrin}{Chen \&
  Guestrin}{2016}]{Chen2016}
Chen T.,  Guestrin C.,  2016, in 22nd ACM sigkdd international conference on
  knowledge discovery and data mining. ACM, pp 785 -- 794,
  \mn@doi{10.1145/2939672.2939785}

\bibitem[\protect\citeauthoryear{Creevey et~al.,}{Creevey
  et~al.}{2017}]{Creevey2017}
Creevey O.,  et~al., 2017, \mn@doi [\aap] {10.1051/0004-6361/201629496}, 601, 1

\bibitem[\protect\citeauthoryear{Debosscher et~al.,}{Debosscher
  et~al.}{2009}]{Debosscher2009}
Debosscher J.,  et~al., 2009, \mn@doi [\aap] {10.1051/0004-6361/200911618},
  506, 519

\bibitem[\protect\citeauthoryear{Debosscher, Blomme, Aerts  \&
  De~Ridder}{Debosscher et~al.}{2011}]{Debosscher2011}
Debosscher J.,  Blomme J.,  Aerts C.,   De~Ridder J.,  2011, \mn@doi [\aap]
  {10.1051/0004-6361/201015647}, 529, A89

\bibitem[\protect\citeauthoryear{Dietterich}{Dietterich}{2000}]{Dietterich2000}
Dietterich T.~G.,  2000, in International workshop on multiple classifier
  systems. pp 1--15, \mn@doi{https://doi.org/10.1007/3-540-45014-9{\_}1}, \url
  {http://www.cs.orst.edu/~tgd
  http://citeseerx.ist.psu.edu/viewdoc/summary?doi=10.1.1.34.4718
  https://link.springer.com/chapter/10.1007/3-540-45014-9_1}

\bibitem[\protect\citeauthoryear{Djorgovski, Mahabal, Drake, Graham  \&
  Donalek}{Djorgovski et~al.}{2013}]{Djorgovski2013}
Djorgovski G.~S.,  Mahabal A.~A.,  Drake A.,  Graham M.~J.,   Donalek C.,
  2013, in Oswalt T.,  Bond H.~E.,  eds, , Planets, Stars and Stellar Systems
  Volume 2: Astronomical Techniques, Software, and Data.
Springer, Dordrecht, Chapt.~5, pp 223--281,
  \mn@doi{10.1007/978-94-007-5618-2{\_}5}, \url
  {http://link.springer.com/10.1007/978-94-007-5618-2_5}

\bibitem[\protect\citeauthoryear{Donalek et~al.,}{Donalek
  et~al.}{2013}]{Donalek2013}
Donalek C.,  et~al., 2013, in Proceedings - 2013 IEEE International Conference
  on Big Data, Big Data 2013. IEEE, pp 35--41,
  \mn@doi{10.1109/BigData.2013.6691731}, \url
  {http://ieeexplore.ieee.org/document/6691731/}

\bibitem[\protect\citeauthoryear{Eyer \& Blake}{Eyer \& Blake}{2005}]{Eyer2005}
Eyer L.,  Blake C.,  2005, \mn@doi [\mnras] {10.1111/j.1365-2966.2005.08651.x},
  358, 30

\bibitem[\protect\citeauthoryear{Fawcett}{Fawcett}{2006}]{Fawcett2006}
Fawcett T.,  2006, \mn@doi [Pattern Recognition Letters]
  {10.1016/j.patrec.2005.10.010}, 27, 861

\bibitem[\protect\citeauthoryear{Feeney et~al.,}{Feeney
  et~al.}{2005}]{Feeney2005}
Feeney S.,  et~al., 2005, \mn@doi [\aj] {10.1086/430844}, 130, 28

\bibitem[\protect\citeauthoryear{Freund}{Freund}{1995}]{Freund1995}
Freund Y.,  1995, {Boosting a weak learning algorithm by majority},
  \mn@doi{10.1006/inco.1995.1136}

\bibitem[\protect\citeauthoryear{Friedman}{Friedman}{1984}]{friedman1984}
Friedman J.~H.,  1984, Technical report, {A variable span smoother}.
Standford Univ CA Laboratory for Computational Statistics

\bibitem[\protect\citeauthoryear{Friedman}{Friedman}{2001}]{Freidman2001}
Friedman J.~H.,  2001, \mn@doi [The Annals of Statistics]
  {10.1017/CBO9781107415324.004}, 29, 1189

\bibitem[\protect\citeauthoryear{Friedman}{Friedman}{2002}]{Friedman2002}
Friedman J.~H.,  2002, \mn@doi [Computational Statistics and Data Analysis]
  {10.1016/S0167-9473(01)00065-2}, 38, 367

\bibitem[\protect\citeauthoryear{Gilliland et~al.,}{Gilliland
  et~al.}{2010}]{Gilliland2010}
Gilliland R.~L.,  et~al., 2010, \mn@doi [\pasp] {10.1086/650399}, 122, 131

\bibitem[\protect\citeauthoryear{Graczyk et~al.,}{Graczyk
  et~al.}{2011}]{Graczyk2011}
Graczyk D.,  et~al., 2011, \actaa, 61, 103

\bibitem[\protect\citeauthoryear{Graham, Drake, Dimitrov, Mahabal, Donalek,
  Duan  \& Maker}{Graham et~al.}{2013}]{Graham2013}
Graham M.~J.,  Drake A.~J.,  Dimitrov D.,  Mahabal A.~A.,  Donalek C.,  Duan
  V.,   Maker A.,  2013, \mn@doi [\mnras] {10.1093/mnras/stt1264}, 434, 3423

\bibitem[\protect\citeauthoryear{Granitto, Furlanello, Biasioli  \&
  Gasperi}{Granitto et~al.}{2006}]{Granitto2006}
Granitto P.~M.,  Furlanello C.,  Biasioli F.,   Gasperi F.,  2006, \mn@doi
  [Chemometrics and Intelligent Laboratory Systems]
  {10.1016/j.chemolab.2006.01.007}, 83, 83

\bibitem[\protect\citeauthoryear{Grigahc{\`{e}}ne et~al.,}{Grigahc{\`{e}}ne
  et~al.}{2010}]{Grigahcene2010}
Grigahc{\`{e}}ne A.,  et~al., 2010, \mn@doi [\apjl]
  {10.1088/2041-8205/713/2/L192}, 713, 192

\bibitem[\protect\citeauthoryear{Hon, Stello  \& Yu}{Hon
  et~al.}{2017}]{Hon2017}
Hon M.,  Stello D.,   Yu J.,  2017, \mn@doi [\mnras] {10.1093/mnras/stx1174},
  469, 4578

\bibitem[\protect\citeauthoryear{Hon, Stello  \& Yu}{Hon
  et~al.}{2018a}]{Hon2018}
Hon M.,  Stello D.,   Yu J.,  2018a, \mn@doi [\mnras] {10.1093/MNRAS/STY483},
  476, 3233

\bibitem[\protect\citeauthoryear{{Hon}, {Stello}  \& {Zinn}}{{Hon}
  et~al.}{2018b}]{Hon2018b}
{Hon} M.,  {Stello} D.,   {Zinn} J.~C.,  2018b, \mn@doi [\apj]
  {10.3847/1538-4357/aabfdb}, \href
  {https://ui.adsabs.harvard.edu/abs/2018ApJ...859...64H} {859, 64}

\bibitem[\protect\citeauthoryear{Howell et~al.,}{Howell
  et~al.}{2014}]{Howell2014}
Howell S.~B.,  et~al., 2014, \mn@doi [\pasp] {10.1086/676406}, 126, 398

\bibitem[\protect\citeauthoryear{Jayasinghe et~al.,}{Jayasinghe
  et~al.}{2018a}]{Jayasinghe2018a}
Jayasinghe T.,  et~al., 2018a, \mn@doi [\mnras] {10.1093/mnras/sty838}, 477,
  3145

\bibitem[\protect\citeauthoryear{Jayasinghe et~al.,}{Jayasinghe
  et~al.}{2018b}]{Jayasinghe2018}
Jayasinghe T.,  et~al., 2018b, eprint, arXiv:1809.07329

\bibitem[\protect\citeauthoryear{Kim \& Bailer-Jones}{Kim \&
  Bailer-Jones}{2016}]{Kim2016}
Kim D.-W.,  Bailer-Jones C. A.~L.,  2016, \mn@doi [\aap]
  {10.1051/0004-6361/201527188}, 587, A18

\bibitem[\protect\citeauthoryear{Kim, Protopapas, Bailer-Jones, Byun, Chang,
  Marquette  \& Shin}{Kim et~al.}{2014}]{Kim2014}
Kim D.-W.,  Protopapas P.,  Bailer-Jones C. A.~L.,  Byun Y.-I.,  Chang S.-W.,
  Marquette J.-B.,   Shin M.-S.,  2014, \mn@doi [\aap]
  {10.1051/0004-6361/201323252}, 566, A43

\bibitem[\protect\citeauthoryear{Kirk et~al.,}{Kirk et~al.}{2016}]{Kirk2016}
Kirk B.,  et~al., 2016, \mn@doi [\aj] {10.3847/0004-6256/151/3/68}, 151, 68

\bibitem[\protect\citeauthoryear{Koch et~al.,}{Koch et~al.}{2010}]{Koch2010}
Koch D.~G.,  et~al., 2010, \mn@doi [\apjl] {10.1088/2041-8205/713/2/L79}, 713,
  L79

\bibitem[\protect\citeauthoryear{Li, Ray  \& Lindsay}{Li et~al.}{2007}]{Li2007}
Li J.,  Ray S.,   Lindsay B.~G.,  2007, Journal of Machine Learning Research,
  8, 1687

\bibitem[\protect\citeauthoryear{Lin, Williamson, Borne  \& DeBarr}{Lin
  et~al.}{2012}]{Lin2012}
Lin J.,  Williamson S.,  Borne K.,   DeBarr D.,  2012, in Ali K.~M.,
  Srivastava A.~N.,  Way M.~J.,   Scargle J.,  eds, , Advances in machine
  learning and data mining for astronomy.
Chapman and Hall/CRC, Chapt.~28, pp 617 -- 639

\bibitem[\protect\citeauthoryear{Lomb}{Lomb}{1976}]{Lomb1976}
Lomb N.~R.,  1976, Astrophysics and Space Science, 39, 447

\bibitem[\protect\citeauthoryear{Lundberg \& Lee}{Lundberg \&
  Lee}{2017}]{Lundberg2017}
Lundberg S.~M.,  Lee S.-i.,  2017, in Advances in Neural Information Processing
  Systems. pp 4765 -- 4774

\bibitem[\protect\citeauthoryear{Lundberg, Cs, Edu, Cs  \& Edu}{Lundberg
  et~al.}{2017}]{Lundberg2017a}
Lundberg S.~M.,  Cs S.,  Edu W.,  Cs S.,   Edu W.,  2017, eprint,
  arXiv:1706.06060

\bibitem[\protect\citeauthoryear{Mairal \& Yu}{Mairal \& Yu}{2012}]{Mairal2012}
Mairal J.,  Yu B.,  2012, preprint, arXiv:1205.0079

\bibitem[\protect\citeauthoryear{Martins, Lopes  \& Casanellas}{Martins
  et~al.}{2017}]{Martins2017}
Martins A.,  Lopes I.,   Casanellas J.,  2017, \mn@doi [\prd]
  {10.1103/PhysRevD.95.023507}, 95, 1

\bibitem[\protect\citeauthoryear{Mathur et~al.,}{Mathur
  et~al.}{2012}]{Mathur2012}
Mathur S.,  et~al., 2012, \mn@doi [\apj] {10.1088/0004-637X/749/2/152}, 749

\bibitem[\protect\citeauthoryear{Matijevi{\v{c}}, Pr{\v{s}}a, Orosz, Welsh,
  Bloemen  \& Barclay}{Matijevi{\v{c}} et~al.}{2012}]{Matijevic2012}
Matijevi{\v{c}} G.,  Pr{\v{s}}a A.,  Orosz J.~A.,  Welsh W.~F.,  Bloemen S.,
  Barclay T.,  2012, \mn@doi [\aj] {10.1088/0004-6256/143/5/123}, 143

\bibitem[\protect\citeauthoryear{McNamara, Jackiewicz  \& McKeever}{McNamara
  et~al.}{2012}]{McNamara2012}
McNamara B.~J.,  Jackiewicz J.,   McKeever J.,  2012, \mn@doi [\aj]
  {10.1088/0004-6256/143/4/101}, 143

\bibitem[\protect\citeauthoryear{Metcalfe et~al.,}{Metcalfe
  et~al.}{2014}]{Metcalfe2014}
Metcalfe T.~S.,  et~al., 2014, \mn@doi [\apjsupp] {10.1088/0067-0049/214/2/27},
  214

\bibitem[\protect\citeauthoryear{Michel et~al.,}{Michel
  et~al.}{2008}]{Michel2008}
Michel E.,  et~al., 2008, Communications in Asteroseismology, 156, 73

\bibitem[\protect\citeauthoryear{Natekin \& Knoll}{Natekin \&
  Knoll}{2013}]{Natekin2013}
Natekin A.,  Knoll A.,  2013, \mn@doi [Frontiers in Neurorobotics]
  {10.3389/fnbot.2013.00021}, 7

\bibitem[\protect\citeauthoryear{Nemec, Cohen, Ripepi, Derekas, Moskalik,
  Sesar, Chadid  \& Bruntt}{Nemec et~al.}{2013}]{Nemec2013MetalField}
Nemec J.~M.,  Cohen J.~G.,  Ripepi V.,  Derekas A.,  Moskalik P.,  Sesar B.,
  Chadid M.,   Bruntt H.,  2013, \mn@doi [\apj] {10.1088/0004-637X/773/2/181},
  773

\bibitem[\protect\citeauthoryear{Nun, Protopapas, Sim, Zhu, Dave, Castro  \&
  Pichara}{Nun et~al.}{2015}]{Nun2015}
Nun I.,  Protopapas P.,  Sim B.,  Zhu M.,  Dave R.,  Castro N.,   Pichara K.,
  2015, eprints, arXiv:1506.00010

\bibitem[\protect\citeauthoryear{Oza}{Oza}{2012}]{Oza2012}
Oza N.,  2012, in , Advances in Machine Learning and Data Mining for Astronomy.
Chapman and Hall/CRC, Chapt.~23, pp 505 -- 522

\bibitem[\protect\citeauthoryear{Pawlak et~al.,}{Pawlak
  et~al.}{2016}]{Pawlak2016}
Pawlak M.,  et~al., 2016, \actaa, 66, 421

\bibitem[\protect\citeauthoryear{Perrone \& Cooper}{Perrone \&
  Cooper}{1993}]{Perrone1993}
Perrone M.~P.,  Cooper L.~N.,  1993, in Mammone R.~J.,  ed., , Artficial Neural
  Networks for Speech and Vission.
Chapman and Hall/CRC, London, pp 126 -- 142

\bibitem[\protect\citeauthoryear{Poleski et~al.,}{Poleski
  et~al.}{2010}]{Poleski2010}
Poleski R.,  et~al., 2010, \actaa, 60, 1

\bibitem[\protect\citeauthoryear{Pr{\v{s}a} et~al.,}{Pr{\v{s}a}
  et~al.}{2011}]{Prsa2011}
Pr{\v{s}a} A.,  et~al., 2011, \mn@doi [\aj] {10.1088/0004-6256/141/3/83}, 141

\bibitem[\protect\citeauthoryear{{Rauer} et~al.,}{{Rauer}
  et~al.}{2014}]{Rauer2014}
{Rauer} H.,  et~al., 2014, \mn@doi [Experimental Astronomy]
  {10.1007/s10686-014-9383-4}, \href
  {https://ui.adsabs.harvard.edu/abs/2014ExA....38..249R} {38, 249}

\bibitem[\protect\citeauthoryear{Reinhold, Reiners  \& Basri}{Reinhold
  et~al.}{2013}]{Reinhold2013}
Reinhold T.,  Reiners A.,   Basri G.,  2013, \mn@doi [\aap]
  {10.1051/0004-6361/201321970}, 560, A4

\bibitem[\protect\citeauthoryear{Richards et~al.,}{Richards
  et~al.}{2011}]{Richards2011}
Richards J.~W.,  et~al., 2011, \mn@doi [\apj] {10.1088/0004-637X/733/1/10}, 733

\bibitem[\protect\citeauthoryear{Ricker et~al.,}{Ricker
  et~al.}{2014}]{Ricker2014}
Ricker G.~R.,  et~al., 2014, \mn@doi [Journal of Astronomical Telescopes,
  Instruments, and Systems] {10.1117/1.JATIS.1.1.014003}, 1, 014003

\bibitem[\protect\citeauthoryear{Rucinski, Carroll, Kuschnig, Matthews  \&
  Stibrany}{Rucinski et~al.}{2003}]{Rucinski2003}
Rucinski S.,  Carroll K.,  Kuschnig R.,  Matthews J.,   Stibrany P.,  2003,
  \mn@doi [Advances in Space Research] {10.1016/S0273-1177(02)00628-2}, 31, 371

\bibitem[\protect\citeauthoryear{Saito \& Rehmsmeier}{Saito \&
  Rehmsmeier}{2015}]{Saito2015}
Saito T.,  Rehmsmeier M.,  2015, \mn@doi [PLoS One]
  {10.1371/journal.pone.0118432}, 10, 1

\bibitem[\protect\citeauthoryear{Sarro, S{\'{a}}nchez-Fern{\'{a}}ndez  \&
  Gim{\'{e}}nez}{Sarro et~al.}{2006}]{Sarro2006}
Sarro L.~M.,  S{\'{a}}nchez-Fern{\'{a}}ndez C.,   Gim{\'{e}}nez Ã.,  2006,
  \mn@doi [\aap] {10.1051/0004-6361:20052830}, 446, 395

\bibitem[\protect\citeauthoryear{Sarro, Debosscher, Aerts  \&
  L{\'{o}}pez}{Sarro et~al.}{2009}]{Sarro2009b}
Sarro L.~M.,  Debosscher J.,  Aerts C.,   L{\'{o}}pez M.,  2009, \mn@doi [\aap]
  {10.1051/0004-6361/200912009}, 506, 535

\bibitem[\protect\citeauthoryear{Sarro et~al.,}{Sarro et~al.}{2013}]{Sarro2013}
Sarro L.~M.,  et~al., 2013, \mn@doi [\aap] {10.1051/00046361/201220184}, 550,
  A120

\bibitem[\protect\citeauthoryear{Scargle}{Scargle}{1982}]{Scargle1982}
Scargle J.~D.,  1982, \mn@doi [\apj] {10.1086/160554}, 263, 835

\bibitem[\protect\citeauthoryear{Silla \& Freitas}{Silla \&
  Freitas}{2011}]{Silla2011}
Silla C.~N.,  Freitas A.~A.,  2011, \mn@doi [Data Mining and Knowledge
  Discovery] {10.1007/s10618-010-0175-9}, 22, 31

\bibitem[\protect\citeauthoryear{Smith et~al.,}{Smith et~al.}{2012}]{Smith2012}
Smith J.~C.,  et~al., 2012, \mn@doi [\pasp] {10.1080/00207543.2016.1229067},
  124, 1000

\bibitem[\protect\citeauthoryear{Soszy{\'{n}}ski, Udalski, Szyma{\'{n}}ski,
  Pietrzy{\'{n}}ski, Wyrzykowski, Szewczyk, Ulaczyk  \&
  Poleski}{Soszy{\'{n}}ski et~al.}{2009}]{Soszynski2009}
Soszy{\'{n}}ski I.,  Udalski A.,  Szyma{\'{n}}ski M.,  Pietrzy{\'{n}}ski G.,
  Wyrzykowski Å.,  Szewczyk O.,  Ulaczyk K.,   Poleski R.,  2009, \actaa, 59,
  239

\bibitem[\protect\citeauthoryear{Soszy{\'{n}}ski et~al.,}{Soszy{\'{n}}ski
  et~al.}{2015}]{Soszynski2015a}
Soszy{\'{n}}ski I.,  et~al., 2015, \actaa, 65, 1

\bibitem[\protect\citeauthoryear{Soszy{\'{n}}ski et~al.,}{Soszy{\'{n}}ski
  et~al.}{2016}]{Soszynski2016}
Soszy{\'{n}}ski I.,  et~al., 2016, \actaa, 66, 131

\bibitem[\protect\citeauthoryear{Stello et~al.,}{Stello
  et~al.}{2013}]{Stello2013}
Stello D.,  et~al., 2013, \mn@doi [\apjl] {10.1088/2041-8205/765/2/L41}, 765

\bibitem[\protect\citeauthoryear{Stumpe et~al.,}{Stumpe
  et~al.}{2012}]{Stumpe2012}
Stumpe M.~C.,  et~al., 2012, \mn@doi [\pasp] {10.1086/667698}, 124, 985

\bibitem[\protect\citeauthoryear{Udalski et~al.,}{Udalski
  et~al.}{1994}]{Udalski1994}
Udalski A.,  et~al., 1994, \apj, 426, 69

\bibitem[\protect\citeauthoryear{Udalski, Szyma{\'{n}}ski, Soszy{\'{n}}ski  \&
  Poleski}{Udalski et~al.}{2008}]{Udalski2008}
Udalski A.,  Szyma{\'{n}}ski M.~K.,  Soszy{\'{n}}ski I.,   Poleski R.,  2008,
  \actaa, 58, 69

\bibitem[\protect\citeauthoryear{Udalski, Szymanski  \& Szymanski}{Udalski
  et~al.}{2015}]{Udalski2015}
Udalski A.,  Szymanski M.~K.,   Szymanski G.,  2015, \actaa, 65, 1

\bibitem[\protect\citeauthoryear{Uytterhoeven et~al.,}{Uytterhoeven
  et~al.}{2011}]{Uytterhoeven2011}
Uytterhoeven K.,  et~al., 2011, \mn@doi [\aap] {10.1051/0004-6361/201117368},
  534, 1

\bibitem[\protect\citeauthoryear{Van~Reeth et~al.,}{Van~Reeth
  et~al.}{2015}]{VanReeth2015}
Van~Reeth T.,  et~al., 2015, \mn@doi [\apjs] {10.1088/0067-0049/218/2/27}, 218,
  27

\bibitem[\protect\citeauthoryear{VanderPlas}{VanderPlas}{2016}]{Vanderplas2016}
VanderPlas J.,  2016, Astrophysics Source Code Library

\bibitem[\protect\citeauthoryear{VanderPlas \& Ivezi{\'{c}}}{VanderPlas \&
  Ivezi{\'{c}}}{2015}]{Vanderplas2015}
VanderPlas J.~T.,  Ivezi{\'{c}} {\v{Z}}.,  2015, \mn@doi [Astrophysical
  Journal] {10.1088/0004-637X/812/1/18}, 812, 1

\bibitem[\protect\citeauthoryear{Wagstaff}{Wagstaff}{2012}]{Wagstaff2012}
Wagstaff K.~L.,  2012, in Ali K.~M.,  Srivastava A.~N.,  Way M.~J.,   Scargle
  J.,  eds, , Advances in Machine Learning and Data Mining for Astronomy.
Chapman and Hall/CRC, Chapt.~25, pp 543 -- 559

\bibitem[\protect\citeauthoryear{Warton}{Warton}{2008}]{Warton2008}
Warton D.~I.,  2008, \mn@doi [Journal of the American Statistical Association]
  {10.1198/016214508000000021}, 103, 340

\makeatother
\end{thebibliography}



\bsp	
\label{lastpage}
\end{document}